\documentstyle[psfig,epsfig,12pt]{article}

\def\myref#1{{\tt #1}}

\def\simge{\mathrel{%
   \rlap{\raise 0.511ex \hbox{$>$}}{\lower 0.511ex \hbox{$\sim$}}}}
\def\simle{\mathrel{
   \rlap{\raise 0.511ex \hbox{$<$}}{\lower 0.511ex \hbox{$\sim$}}}}
 
\def\slashchar#1{\setbox0=\hbox{$#1$}           % set a box for #1
   \dimen0=\wd0                                 % and get its size
   \setbox1=\hbox{/} \dimen1=\wd1               % get size of /
   \ifdim\dimen0>\dimen1                        % #1 is bigger
      \rlap{\hbox to \dimen0{\hfil/\hfil}}      % so center / in box
      #1                                        % and print #1
   \else                                        % / is bigger
      \rlap{\hbox to \dimen1{\hfil$#1$\hfil}}   % so center #1
      /                                         % and print /
   \fi}                                         %
\def\nn{\nonumber}
\def\ts{\thinspace}
\def\tx{\textstyle}
\def\ra{\rightarrow}

\def\ol{\bar}
\def\dag{\dagger}

\def\be{\begin{equation}} 
\def\ee{\end{equation}} 
\def\bea{\begin{eqnarray}}
\def\eea{\end{eqnarray}}
\def\ba{\begin{array}}
\def\ea{\end{array}}
\def\chipr{\chi^{\ts \prime}}
\def\CA{{\cal A}}
\def\CB{{\cal B}}
\def\CC{{\cal C}}
\def\CD{{\cal D}}

\def\CG{{\cal G}}

\def\CL{{\cal L}}
\def\CM{{\cal M}}
\def\CN{{\cal N}}
\def\CO{{\cal O}}

\def\CR{{\cal R}}

\def\ecm{\sqrt{s}}
\def\shat{\hat s}
\def\that{\hat t}
\def\uhat{\hat u}
\def\rshat{\sqrt{\shat}}

\def\atc{\alpha_{TC}}

\def\atro{\alpha_{\tro}}

\def\Ntc{N_{TC}}
\def\suc{SU(3)_C}

\def\sutc{SU(\Ntc)}

\def\uone{U(1)_1}
\def\utwo{U(1)_2}

\def\suone{SU(3)_1}
\def\sutwo{SU(3)_2}
\def\thw{\theta_W}
\def\thc{\theta_3}
\def\kslash{\raise.15ex\hbox{/}\kern-.57em k}

\def\Dggs{\Delta_{gg}(\shat)}
\def\Dggt{\Delta_{gg}(\that)}
\def\Dggu{\Delta_{gg}(\uhat)}
\def\Dggss{\Delta^*_{gg}(\shat)}
\def\Dggts{\Delta^*_{gg}(\that)}

\def\tro{\rho_{T}}
 
\def\troct{\rho_{T8}} 
\def\troctaa{\rho_{11}} 
\def\troctbb{\rho_{22}} 
\def\troctab{\rho_{12}} 
\def\troctabp{\rho_{12'}} 
\def\tropm{\rho_{T}^\pm}
\def\trop{\rho_{T}^+}
\def\trom{\rho_{T}^-}
\def\troz{\rho_{T}^0}
\def\tom{\omega_T}
\def\tpi{\pi_T}
\def\tpipm{\pi_T^\pm}
\def\tpimp{\pi_T^\mp}
\def\tpip{\pi_T^+}
\def\tpim{\pi_T^-}
\def\tpiz{\pi_T^0}
\def\tpipr{\pi_T^{0 \prime}}

\def\tpioct{\pi_{T8}}

\def\Mv{M_{V_8}}

\def\mev{{\rm MeV}}
\def\gev{{\rm GeV}}
\def\tev{{\rm TeV}}

\def\ipb{{\rm pb}^{-1}}

\def\half{{\textstyle{ { 1\over { 2 } }}}}
\def\third{{\textstyle{ { 1\over { 3 } }}}}
\def\fourth{{\textstyle{ { 1\over { 4 } }}}}

\def\fourthirds{{\textstyle{ { 4\over { 3 } }}}}

\def\nin{\noindent}

\begin{document}
\title{
\vskip -15mm
\begin{flushright}
\vskip -15mm
{\small BUHEP-02-32\\
FERMILAB-Pub-02/267-T\\
hep-ph/0210299\\}
\vskip 5mm
\end{flushright}
{\Large{\bf The Collider Phenomenology of Technihadrons in the \\
Technicolor Straw Man Model}}\\
}
\author{
{\large Kenneth Lane$^{1,2}$\thanks{lane@physics.bu.edu} $\ts$ and Stephen
    Mrenna$^{2}$\thanks{mrenna@fnal.gov}}\\
{\small {$^1$}Department of Physics, Boston University}\\
{\small 590 Commonwealth Avenue, Boston, Massachusetts 02215}\\
{\small {$^2$}Fermi National Accelerator Laboratory}\\
{\small P.O.~Box 500, Batavia, Illinois 60510}\\
}
\maketitle

\begin{abstract}
We discuss the phenomenology of the lightest $SU(3)_C$ singlet and
non-singlet technihadrons in the Straw Man Model of low-scale technicolor
(TCSM). The technihadrons are assumed to be those arising in
topcolor--assisted technicolor models in which topcolor is broken by
technifermion condensates. We improve upon the description of the
color--singlet sector presented in our earlier paper introducing the TCSM
(hep-ph/9903369). These improvements are most important for subprocess
energies well below the masses of the $\tro$ and $\tom$ vector technihadrons
and, therefore, apply especially to $e^+e^-$ colliders such as LEP and a
low--energy linear collider. In the color--octet sector, we consider mixing
of the gluon, the coloron $V_8$ from topcolor breaking, and four isosinglet
color--octet technirho mesons $\troct$. We assume, as expected in walking
technicolor, that these $\troct$ decay into $\ol q q$, $gg$, and $g \tpi$
final states, but not into $\tpi\tpi$, where $\tpi$ is a technipion. All the
TCSM production and decay processes discussed here are included in the event
generator {\sc Pythia}. We present several simulations appropriate for the
Tevatron Collider, and suggest benchmark model lines for further experimental
investigation.
\end{abstract}

%%%%%%%%%%%%%%%%%%%%%%%%%%%%%%%%%
%%%%%%%%%%%%%%%%%%%%%%%%%%%%%%%%%

\newpage

\section*{1. Overview of the TCSM}

In this paper we improve and extend the ``technicolor straw man model''
(TCSM)~\cite{tcsm} of collider signatures for dynamical electroweak and
flavor symmetry breaking. The TCSM is a simple phenomenology of the
lowest--lying vector and pseudoscalar technihadrons expected in technicolor
theories which have both a walking gauge coupling $\atc$ and topcolor
interactions to generate the large mass of the top quark.

The main improvements arise in the color $\suc$ singlet sector of the
model. The treatment in Ref.~\cite{tcsm} assumed fermion--antifermion
annihilations to technihadrons occurred at subprocess cm energy $\rshat \sim
M_{\tro,\ts \tom}$. This is is appropriate in a hadron collider where the
parton $\rshat$ sweeps over the narrow TCSM resonances and is, therefore,
strongly dominated by the poles. In this paper we modify the production
amplitudes to make them accurate for $\rshat$ below the resonances as
well. These modifications involve adding the continuum amplitudes including,
where appropriate, the three--point anomaly amplitudes for production of an
electroweak gauge boson and a technipion~\cite{ehs,Lane:2002wb}. This is
particularly appropriate for technihadron searches at $e^+e^-$ colliders such
as LEP~\cite{Mrenna:1999ks,l3,delphi,opal} (and the proposed linear collider
if its early stage has $\ecm \simle 300\,\gev$).

Our extension of the TCSM includes several low--lying color--non-singlet,
isosinglet states---four vector $\troct$'s, the massive topcolor gauge boson,
$V_8$, and a neutral pseudoscalar $\tpioct$. These particles are most readily
produced and detected in hadron colliders---the Tevatron and the Large Hadron
Collider (LHC)---because of their relatively large coupling to gluons. We have
incorporated the improved and extended production processes in the event
generator {\sc Pythia}~\cite{pythia} to allow detailed investigation by 
collider experiments.

The modern description of technicolor~\cite{tc,etc} requires a walking
technicolor gauge coupling~\cite{wtc} to evade unwanted flavor-changing
neutral currents and the assistance of topcolor interactions that are strong
near 1~TeV~\cite{topcref,tctwohill} to provide a large top quark mass. Both
these elaborations on the basic technicolor/extended technicolor (TC/ETC)
proposal require a large number $N_D$ of technifermion doublets. Many
technifermions are needed to make the beta function of walking technicolor
small. Many also seem required in topcolor-assisted technicolor (TC2) to
generate the hard masses of quarks and leptons, to induce the correct mixing
between heavy and light quarks, and to break topcolor symmetry down to
ordinary color (via technifermion condensation). A large number of
technidoublets implies a small technipion decay constant $F_T \simeq
F_\pi/\sqrt{N_D}$, where $F_\pi = 246\,\gev$. This, in turn, implies a
technicolor scale $\Lambda_{TC} \sim F_T$ and masses of a few hundred GeV for
the lowest--lying technihadrons ($\tpi$, $\tro$,
$\tom$)~\cite{multi,elw}. The signatures of this low--scale technicolor have
been sought in Run~I of the Tevatron
Collider~\cite{CDFWtpi,CDFgtpi,CDFjets,Hoffman:mk,CDFLQ}. In the past two
years, several LEP experiments have published limits on color--singlet
technihadrons~\cite{l3,delphi,opal} which in some cases exceed those set at
Fermilab. Run~II will significantly extend the reach of Run~I in
color--singlet and octet sectors. If these low--scale technihadrons exist,
they certainly will be seen in LHC experiments~\cite{ATLAS}.

The TCSM provides a simple framework for these searches. First, and probably
most important, we assume that the lowest-lying bound states of the lightest
technifermions can be considered {\it in isolation}. The lightest
technifermions are expected to be an isodoublet of color singlets, $(T_U,
T_D)$. Color triplets, discussed below, will be heavier because of $\suc$
contributions to their hard (chiral symmetry breaking) masses. We assume that
all technifermions transform under technicolor $SU(\Ntc)$ as fundamentals.
The electric charges of $(T_U,T_D)$ are $Q_U$ and $Q_D = Q_U - 1$. The
color--singlet bound states we consider are vector and pseudoscalar
mesons. The vectors include a spin-one isotriplet $\tro^{\pm,0}$ and an
isosinglet $\tom$. In topcolor-assisted technicolor, there is no need to
invoke large isospin-violating extended technicolor interactions to explain
the top-bottom splitting. Thus, techni-isospin can be a good approximate
symmetry, so that $\tro$ and $\tom$ are nearly degenerate. Their mixing will
be described in the neutral--sector propagator matrix, $\Delta_0$, in
Eq.~(\ref{eq:gzprop}) below. The new, improved formulas for the decays and
production of these technivector mesons will be presented in Sections~2
and~3.

The lightest pseudoscalar $(T_U,T_D)$ bound states, color--singlet
technipions, also comprise an isotriplet $\Pi_T^{\pm,0}$ and an isosinglet
$\Pi_T^{0 \prime}$. However, these are not mass eigenstates. Our second
important assumption for the TCSM is that the isovectors are simple {\it
two-state mixtures} of the longitudinal weak bosons $W_L^\pm$, $Z_L^0$---the
true Goldstone bosons of dynamical electroweak symmetry breaking in the limit
that the $SU(2) \otimes U(1)$ couplings $g,g'$ vanish---and mass-eigenstate
pseudo-Goldstone technipions $\tpi^\pm, \tpiz$:
\be\label{eq:pistates}
 \vert\Pi_T\rangle = \sin\chi \ts \vert
W_L\rangle + \cos\chi \ts \vert\tpi\rangle\ts.
\ee
We assume that $\sutc$ gauge interactions dominate the binding of all
technifermions into technihadrons. Then the decay constants of color--singlet
and nonsinglet $\tpi$ are approximately equal, $F_T \simeq F_\pi/\sqrt{N_D}$,
and the mixing angle is given by
\be\label{eq:sinchi}
\sin\chi \simeq F_T/F_\pi \simeq 1/\sqrt{N_D} \ll 1 \ts.
\ee

Similarly, $\vert\Pi_T^{0 \prime} \rangle = \cos\chipr \ts
\vert\tpipr\rangle\ + \cdots$, where $\chipr$ is another mixing angle and the
ellipsis refer to other technipions needed to eliminate the two-technigluon
anomaly from the $\Pi_T^{0 \prime}$ chiral current. It is unclear whether,
like $\tro$ and $\tom$, these neutral technipions will be degenerate as we
have previously proposed~\cite{elw}. On one hand, they both contain the
lightest $\ol T T$ as constituents. On the other, because of the anomaly
cancellation, $\tpipr$ must contain other heavier technifermions such as the
color triplets we discuss below. If $\tpiz$ and $\tpipr$ are nearly
degenerate and if their widths are roughly equal, there may be appreciable
$\tpiz$--$\tpipr$ mixing and, then, the lightest neutral technipions will be
ideally-mixed $\ol T_U T_U$ and $\ol T_D T_D$ bound states. Searches for
these technipions ought to consider both possibilities: they are nearly
degenerate or not at all.

Color--singlet technipion decays are mediated by ETC and (in the case of
$\tpipr$) $\suc$ interactions. In the TCSM they are taken to
be~\cite{tcsmrates}:
\bea\label{eq:tpiwidths}
 \Gamma(\tpi \ra \ol f' f) &=& {1 \over {16\pi F^2_T}}
 \ts N_f \ts p_f \ts C^2_{1f} (m_f + m_{f'})^2 \nn \\ \nn \\
 \Gamma(\tpipr \ra gg) &=& {1 \over {128 \pi^3 F^2_T}} 
 \ts \alpha_C^2 \ts C^2_{1g} \ts \Ntc^2 \ts M_{\tpipr}^{{3\over{2}}} \ts .
\eea
Like elementary Higgs bosons, technipions are {\it expected} to couple to
fermion mass. Thus, $C_{1f}$ is an ETC-model dependent factor of order one
{\it except} that TC2 suggests $\vert C_{1t}\vert \simle m_b/m_t$ so that
there is not a strong preference for technipions to decay to (or from) top
quarks. The number of colors of fermion~$f$ is $N_f$. The fermion momentum is
$p_f$. The QCD coupling $\alpha_C$ is evaluated at $M_{\tpi}$; and $C^2_{1g}$
is a Clebsch of order one. The default values of these and other parameters
are tabulated at the end of this paper. For $M_{\tpi} < m_t - m_b$, these
technipions are expected to decay as follows: $\tpip \ra c \ol b$ or $c \ol
s$ or even $\tau^+ \nu_\tau$; $\tpiz \ra b \ol b$ and, perhaps $c \ol c$,
$\tau^+\tau^-$; and $\tpipr \ra gg$, $b \ol b$, $c \ol c$, $\tau^+\tau^-$.

The breaking of topcolor to ordinary color is most economically achieved by
color--triplet technifermions~\cite{tctwokl}. Therefore, we expect
$\suc$--nonsinglet technihadrons to exist. The lightest of these may have
masses not very much larger than their color--singlet
counterparts. Production and detection of color--nonsinglet $\tro$ and $\tpi$
have been studied before, both theoretically~\cite{ehlq,multiklrm} and
experimentally~\cite{CDFjets,Hoffman:mk,CDFLQ}. In these studies it was
assumed that there exists one doublet of color--triplet technifermions, one
or more doublets of color--singlets, and their associated technihadrons. In
the likely case that techni--isospin is a good symmetry, the most accessible
states would then be an isosinglet $\troct$, produced as an $s$--channel
resonance in $\ol q q$ and $gg$ collisions, and the technipions (as well as
dijets) into which $\troct$ may decay. With the advent of TC2, this simple
model became obsolete.

In TC2 models, the existence of a large $\ol tt$ (but not $\ol bb$)
condensate and mass is due to $\suone \otimes \uone$ gauge interactions which
are strong near 1~TeV~\cite{tctwohill}. The $\suone$ interaction is $t$--$b$
symmetric while $\uone$ couplings are $t$--$b$ asymmetric. This makes these
forces supercritical for the top quark, but subcritical for bottom. There are
weaker $\sutwo \otimes \utwo$ gauge interactions in which light quarks (and
leptons) may or may not participate.

The two $U(1)$'s must be broken to weak hypercharge $U(1)_Y$ at an energy
somewhat higher than 1~TeV by electroweak--singlet
condensates~\cite{tctwokl}. This breaking results in a heavy color--singlet
$Z'$ boson whose physics we do not consider in this paper (see, however,
Refs.~\cite{ctzprime, trzprime}). The two $SU(3)$'s are broken to their
diagonal $\suc$ subgroup. When this happens, a degenerate octet of
``colorons'', $V_8$, mediate the broken topcolor $SU(3)$ interactions. There
are two variants of TC2: The ``standard'' version~\cite{tctwohill}, in which
only the third generation quarks are $\suone$ triplets, and the
``flavor--universal'' version~\cite{ccs} in which all quarks are $\suone$
triplets. We consider both in this paper.

In either version of TC2, the $SU(3)$'s are economically broken to $\suc$
subgroup by using technicolor and $\uone$ interactions, both strong near
1~TeV. Following Ref.~\cite{tctwokl}, we assume the existence of two
electroweak doublets of technifermions, $T_1= (U_1,D_1)$ and $T_2 =
(U_2,D_2)$, which transform respectively as $({\bf 3},{\bf 1},{\bf \Ntc})$
and $({\bf 1},{\bf 3},{\bf \Ntc})$ under the two color groups and
technicolor. The desired pattern of symmetry breaking occurs if $SU(\Ntc)$
and $\uone$ interactions work together to induce electroweak and $\suone
\otimes \sutwo$ noninvariant condensates (which, of course, preserve $\suc$
and $U(1)_{EM}$) 
\be\label{eq:Tconds}
\langle \ts \ol U_{iL} U_{jR} \rangle = -W^{U\ts *}_{ij} \Delta_T \ts, \quad
\langle \ts \ol D_{iL} D_{jR} \rangle = -W^{D\ts *}_{ij} \Delta_T \ts, \quad 
(i,j = 1,2)\ts.
\ee
Here, $W^U$ and $W^D$ are {\it nondiagonal} $U(2)$ matrices and $\Delta_T
\sim \Lambda^3_{TC}$. Assuming the condensates are parity--conserving,
$(W^{U,D}_{ij})^* = W^{U,D}_{ji}$. Then, unitarity implies
\bea\label{eq:unitarity}
&& W^{U,D}_{11} + W^{U,D}_{22} = 0\ts, \nn\\
&& W^{U,D}_{12} = (W^{U,D}_{21})^* = e^{i\phi_{U,D}}
\sqrt{1-(W^{U,D}_{11})^2} \ts,
\eea
where $\phi_{U,D}$ are phases to be chosen.

This minimal TC2 scenario leads to a rich spectrum of color--nonsinglet
states readily accessible in hadron collisions. We concentrate on the
lowest--lying isosinglet color octets. In addition to the colorons, $V_8$,
these include four $\troct$ formed from $\ol T_i T_j$, and a technipion
$\tpioct$. These states are constructed in Section~4. In Section~5 we discuss
the decay rates of $\troct$ and $V_8$ under the simplifying assumption that
$M_{\rho_{T8}} < 2 M_{\tpi}$. As in the color--singlet TCSM, this is likely
because the large coupling of walking technicolor significantly enhances the
ETC contribution to $M_{\pi_T}$~\cite{multi}. In the extreme walking $\atc$
limit, $M_{\pi_T} \simeq M_{\rho_{T8}}$. In Section~6 we present the dijet
production cross sections, using the fully mixed propagator of gluons,
colorons and the four $\troct$. These cross sections are now encoded in the
{\sc Pythia} event generator~\cite{pythia}. Simulated signals (at the parton
level) appropriate to Run~I of the Tevatron Collider are presented in
Section~7, with special attention to the effects on the $\ol t t$ invariant
mass distribution.

Two comments are in order before moving on. First, it has been argued in
Ref.~\cite{blt} that $B_d$--$\ol B_d$ mixing constrains $\Mv\tan\thc >
1$--$2\,\tev$ in standard, but not flavor--universal, TC2 (also see
Ref.~\cite{ehsbbmix}). Here, $\thc$ is the $\suone$--$\sutwo$ mixing angle
defined in Section~4 below. This constraint relies on the form of quark mass
and mixing matrices expected in ETC/TC2. We view as independent and
complementary the limit on $\Mv\tan\thc$ that may be derived from
collider jet data. Second, Bertram and Simmons have used D\O\ jet data from
Run~I to place the limit $\Mv\tan\thc > 0.84\,\tev$ in flavor--universal
TC2~\cite{ibehs}. Their analysis did not include the potential complicating
effects of $\troct$ and their interference with $V_8$. Our TCSM analysis
closes this loophole. We find limits on the coloron mass as high or higher
than theirs.

\section*{2. Decays of Color--Singlet $\tro$ and $\tom$}

In the limit that the electroweak couplings $g,g' = 0$, the $\tro$ and $\tom$
decay as
\bea\label{eq:vt_decays}
\tro &\ra& \Pi_T \Pi_T = \cos^2 \chi\ts (\tpi\tpi) + 2\sin\chi\ts\cos\chi
\ts (W_L\tpi) + \sin^2 \chi \ts (W_L W_L) \ts; \nn \\\nn\\
\tom &\ra& \Pi_T \Pi_T \Pi_T = \cos^3 \chi \ts (\tpi\tpi\tpi) + \cdots \ts.
\eea
The $\tro$ decay amplitude is
\be\label{eq:rhopipi}
\CM(\tro(q) \ra \pi_A(p_1) \pi_B(p_2)) = g_{\tro} \ts \CC_{AB}
\ts \epsilon(q)\cdot(p_1 - p_2) \ts,
\ee
where $\epsilon(q)$ is the $\tro$ polarization vector; $\atro \equiv
g_{\tro}^2/4\pi = 2.91(3/\Ntc)$ is scaled naively from QCD and $\Ntc = 4$ is
used in calculations; and
\be\label{eq:ccab}
\ba{ll}
\CC_{AB} &= \left\{\ba{ll} \sin^2\chi  & {\rm for} \ts\ts\ts\ts W_L^+ W_L^-
\ts\ts\ts\ts {\rm or} \ts\ts\ts\ts  W_L^\pm Z_L^0 \\
\sin\chi \cos\chi & {\rm for} \ts\ts\ts\ts W_L^+ \tpim, W_L^- \tpip
\ts\ts\ts\ts  {\rm or} \ts\ts\ts\ts W_L^\pm \tpiz, Z_L^0 \tpipm \\
\cos^2\chi & {\rm for} \ts\ts\ts\ts \tpip\tpim  \ts\ts\ts\ts {\rm or}
\ts\ts \ts\ts \tpipm\tpiz \ts.
\ea \right.
\ea
\ee
The $\tro$ decay rate to two technipions is then (for later use in cross
sections, we quote the energy-dependent width for a $\tro$ mass of
$\sqrt{\shat}$)
\be\label{eq:trhopipi}
\Gamma(\troz \ra \pi_A^+ \pi_B^-) = \Gamma(\tropm \ra \pi_A^\pm \pi_B^0) =
{2 \atro \CC^2_{AB}\over{3}} \ts {\ts\ts p^3\over {\shat}} \ts,
\ee
where $p = [(\shat - (M_A+M_B)^2) (\shat - (M_A-M_B)^2)]^\half/2\rshat$
is the $\tpi$ momentum in the $\tro$ rest frame.

Walking technicolor enhancements of technipion masses are likely to close off
the channels $\tro \ra \tpi\tpi$, $\tom \ra \tpi\tpi\tpi$ and even the
isospin--violating $\tom \ra \tpi\tpi$~\cite{multi}. A $\troz$ of mass
200~GeV, say, may then decay to $W_L \tpi$ or $W_L W_L$, but such $\tom$
decays are strongly suppressed (see Eq.~(\ref{eq:isov} below). Therefore, all
$\tom$ decays are electroweak: $\tom \ra \gamma\tpiz$, $Z^0\tpiz$, $W^\pm
\tpimp$, and $\ol f f$. Here, $Z$ and $W$ are transversely
polarized.~\footnote{Strictly speaking, the identification of $W$ and $Z$
decay products as longitudinal or transverse is approximate, becoming exact
in the limit of very large $M_{\tro,\tom}$.}  Furthermore, since we expect
$\sin^2\chi \ll 1$, the electroweak decays of $\tro$ to the transverse gauge
bosons $\gamma,W,Z$ plus a technipion may be competitive with the
open-channel strong decays.

As discussed in Ref.~\cite{tcsm}, the amplitude for $V_T = \tro,\tom$
decay to any transversely polarized electroweak boson $G$ plus a technipion
is given by
\bea\label{eq:genamp}
\CM(V_T(q) \ra G(p_1) \tpi(p_2)) &=& {eV_{V_T G\tpi} \over{M_V}}\ts
\epsilon^{\mu\nu\lambda\rho} \ts \epsilon_\mu(q) \ts \epsilon^*_\nu(p_1) \ts
q_{\lambda} \ts p_{1\rho} \\
 &+& {eA_{V_T G\tpi} \over{M_A}} \biggl(\epsilon(q) \cdot \epsilon^*(p_1)\ts
q\cdot p_1 - \epsilon(q) \cdot p_1\ts \epsilon^*(p_1) \cdot q\biggr) \nn
\ts.
\eea
The first term corresponds to the vector coupling of $G$ to the constituent
technifermions of $V_T$ and $\tpi$ and the second term to its axial-vector
coupling. The technicolor--scale parameter $M_V$ is expected to be of order
several 100~GeV.~\footnote{The corresponding $\rho \ra \gamma\pi$ parameter
in QCD is about $400\,\mev$. A large--$N_C$ argument implies $M_V \simeq
(F_T/f_\pi) \ts 400\,\mev \simeq 350\,\gev$.} The mass parameter $M_A$ is
expected to be comparable to $M_V$. Note that the amplitudes for emission of
longitudinally polarized bosons in Eq.~(\ref{eq:rhopipi}) and transversely
polarized ones in Eq.~(\ref{eq:genamp}) are noninterfering, as they should
be. Adopting a ``valence technifermion'' model for the graphs describing
Eq.~(\ref{eq:genamp})---a model which works very well for $\omega, \rho \ra
\gamma \pi$ and $\gamma \eta$ in QCD---CP-invariance implies that the $V$ and
$A$ coefficients in this amplitude are given in our normalization
by~\footnote{We have neglected decays such as $\troz \ra W_T W_L$ and $\troz
\ra W_T W_T$. The rate for the former is suppressed by $\tan^2 \chi$ relative
to the rate for $\troz \ra W_T \tpi$ while the latter's rate is suppressed by
$\alpha$.}
\be\label{eq:VA}
V_{V_T G\tpi} = 2\ts{\rm Tr}\biggl(Q_{V_T} \{Q^\dagger_{G_V}, \ts
Q^\dagger_{\tpi}\}\biggr) \ts,\qquad
A_{V_T G\tpi} = 2\ts{\rm Tr}\biggl(Q_{V_T} [Q^\dagger_{G_A}, \ts
Q^\dagger_{\tpi}]\biggr) \ts.
\ee
In the TCSM, with electric charges $Q_U$, $Q_D$ for $T_U$, $T_D$, the
generators $Q$ in Eq.~(\ref{eq:VA}) are given by
\bea\label{eq:charges}
Q_{\tom} &=& \left(\ba{cc} \half & 0 \\ 0 & \half \ea\right)\nn\\
Q_{\troz} &=& \left(\ba{cc} \half & 0 \\ 0 & -\half \ea\right)
\ts;\qquad
Q_{\trop} = Q^\dagger_{\trom} =
{1\over{\sqrt{2}}}\left(\ba{cc} 0 & 1 \\ 0 & 0 \ea\right)\nn\\
Q_{\tpiz} &=& \cos\chi \left(\ba{cc} \half & 0 \\ 0 & -\half \ea\right)
\ts;\ts\quad
Q_{\tpip} = Q^\dagger_{\tpim} = {\cos\chi\over{\sqrt{2}}}
 \left(\ba{cc} 0 & 1 \\ 0 & 0  \ea\right)\nn\\
Q_{\tpipr} &=& \cos\chipr \left(\ba{cc} \half & 0 \\ 0
  & \half \ea\right) \nn\\
Q_{\gamma_V} &=& \left(\ba{cc} Q_U& 0 \\ 0 & Q_D \ea\right)
\ts;\qquad
Q_{\gamma_A} = 0 \nn\\
Q_{Z_V} &=& {1\over{\sin\thw \cos\thw}} \left(\ba{cc} \fourth - Q_U
  \sin^2\thw & 0 \\ 0 & -\fourth - Q_D \sin^2\thw \ea\right) \nn\\
Q_{Z_A} &=& {1\over{\sin\thw \cos\thw}} \left(\ba{cc} -\fourth & 0 \\ 0
  &  \fourth \ea\right) \nn\\
Q_{W^+_V} &=& Q^\dagger_{W^-_V} = -Q_{W^+_A} = -Q^\dagger_{W^-_A} = {1\over
  {2\sqrt{2}\sin\thw}}\left(\ba{cc} 0 & 1 \\ 0 & 0 \ea\right) \ts.
\eea

{\flushleft{\begin{table}{
\begin{tabular}{|c|c|c|c|}
\hline
Process & $V_{V_T G\tpi}$ & $A_{V_T G\tpi}$ & $\Gamma(V_T \ra G\tpi)$ \\
\hline\hline
$\tom \ra \gamma \tpiz$& $c_\chi$ & 0 & 0.115 $c^2_\chi$ \\
$\ts\ts\ts\quad \ra \gamma \tpipr$ & $(Q_U + Q_D)\ts c_{\chipr}$ & 0 & 0.320
$c^2_{\chipr}$ \\ 
$\qquad \ra Z^0 \tpiz$ & $c_\chi\cot 2\thw$ & 0 &
$2.9\times 10^{-3}c^2_\chi$ \\ 
$\ts\qquad \ra Z^0 \tpipr$ & $-(Q_U+Q_D)\ts c_{\chipr}\tan \thw$ & 0 &
$5.9\times 10^{-3}c^2_{\chipr}$ \\ 
$\ts\ts\ts\ts\qquad \ra W^\pm \tpimp$ & $c_\chi/(2\sin\thw)$ & 0 &
  $2.4\times 10^{-2}c^2_\chi$ \\ 
\hline
$\troz \ra \gamma \tpiz$ & $(Q_U + Q_D)\ts c_\chi$ & 0 & 0.320 $c^2_\chi$ \\
$\ts\ts\ts\quad \ra \gamma \tpipr$ & $c_{\chipr}$ & 0 & 0.115 $c^2_{\chipr}$\\ 
$\qquad \ra Z^0 \tpiz$ & $-(Q_U+Q_D)\ts c_\chi \tan \thw$ & 0 &
$5.9\times 10^{-3}c^2_\chi$ \\ 
$\ts\qquad \ra Z^0 \tpipr$ & $c_{\chipr}\ts \cot 2\thw$ & 0 &
$2.9 \times 10^{-3}c^2_{\chipr}$ \\ 
$\ts\ts\ts\ts\qquad \ra W^\pm \tpimp$ & 0 & $\pm c_\chi/(2\sin\thw)$ &
0.143 $c^2_\chi$  \\ 
\hline
$\tropm \ra \gamma \tpipm$ & $(Q_U + Q_D)\ts c_\chi$ & 0 & 0.320 $c^2_\chi$\\ 
$\qquad \ra Z^0 \tpipm$ & $-(Q_U+Q_D)\ts c_\chi \tan \thw$ & $\pm c_\chi
\ts /\sin 2\thw$ & 0.153 $c^2_\chi$ \\  
$\ts\ts\ts\qquad \ra W^\pm \tpiz$ & 0 & $\mp c_\chi/(2\sin\thw)$ &
0.143 $c^2_\chi$  \\ 
$\ts\ts\ts\qquad \ra W^\pm \tpipr$ & $c_{\chipr}/(2\sin\thw)$ & 0&
  $2.4\times 10^{-2}c^2_{\chipr}$ \\ 
\hline\hline
\end{tabular}}
\caption{Amplitudes and sample decay rates (in GeV) for $V_T \ra G \tpi$. In
the rate calculations, $M_{V_T} = 210\,\gev$, $M_{\tpi} = 110\,\gev$, $M_V
= M_A = 100\,\gev$; technifermion charges are $Q_U + Q_D
=\textstyle{5\over{3}}$; $c_\chi = \cos\chi$ and $c_{\chipr} = \cos\chipr$;
$G_V$ and $G_A$ refer to decays involving the vector and axial-vector
couplings, respectively. Rates involving both couplings are summed in the
last column.}
\end{table}}}

The decay rate for $V_T \ra G \tpi$ is
\be
\Gamma(V_T \ra G \tpi) = {\alpha V^2_{V_T G\tpi} \ts p^3\over {3M_V^2}} +
{\alpha A^2_{V_T G\tpi} \ts p\ts(3 M_G^2 + 2p^2)\over {6M_A^2}} \ts,
\ee
where $p$ is the $G$-momentum in the $V_T$ rest frame. The $V$ and $A$
coefficients and sample decay rates are listed in Table~1. These are to be
compared with the rates for decay into longitudinal $W$ and $Z$ bosons plus a
technipion. For $M_{\tro} = 210\,\gev$, $M_{\tpi} = 110\,\gev$, and $\Ntc =
4$, they are
\bea\label{eq:longdecay}
\Gamma(\troz \ra W_L^\pm \tpimp) &=& \Gamma(\tropm \ra W_L^\pm \tpiz) =
2.78\sin^2\chi\cos^2\chi \nn \\
\Gamma(\tropm \ra Z_L^0 \tpimp) &=& 0.89\sin^2\chi\cos^2\chi \ts.
\eea
For $\sin^2\chi = 1/9$, our nominal choice, and for $M_V = M_A = 100\,\gev$,
the rates for $\tro$ and $\tom \ra \gamma\tpi$ and for $\tro \ra W_T \tpi$,
$Z_T\tpi$ via axial vector coupling are comparable to these. Obviously, these
transverse boson decay rates fall quickly for greater $M_V$ and $M_A$.

The rate for the isospin-violating decay $\tom \ra W_L^\pm
\tpimp$ can be estimated as
\be\label{eq:isov}
\Gamma(\tom \ra W_L^\pm \tpimp) = \vert\epsilon_{\rho\omega}\vert^2 \ts
\Gamma(\troz \ra W_L^\pm \tpimp) \ts,
\ee
where $\epsilon_{\rho\omega}$ is the $\tro$--$\tom$ mixing amplitude. In QCD,
$\vert \epsilon_{\rho\omega}\vert \simeq 0.05$, so we expect this decay mode
to be entirely negligible.

Finally, for completeness, we record again the decay rates for $\tro, \tom
\ra f \ol f$. The $\tro$ decay rates to fermions with $N_f=1$ or 3~colors
are
\bea\label{eq:trhoff}
\Gamma(\troz \ra f_i \ol f_i) &=& {N_f \ts \alpha^2 p\over
{3 \atro \shat}} \ts \left((\shat - m_i^2)
\ts A_i^0(\shat) + 6 m_i^2\ts \CR e(\CA_{iL}(\shat)
\CA_{iR}^*(\shat))\right) \ts, \nn\\ \\
\Gamma(\trop \ra f_i \ol f'_i) &=& {N_f \ts \alpha^2 p\over
{6 \atro} \shat^2} \ts \left(2\shat^2 - \shat (m_i^2 + m^{'2}_i) -
(m_i^2 - m^{'2}_i)^2\right) A_i^+(\shat) \ts,\nn
\eea
where a unit CKM matrix is assumed in the second equality. The quantities
$A_i$ are given by
\bea\label{eq:afactors}
A_i^\pm(\shat) &=& {1 \over {8\sin^4\thw}} \ts \biggl\vert{\shat \over {\shat
    - \CM^2_W}}\biggr\vert^2 \ts, \nn \\ \nn\\
A_i^0(\shat) &=& \vert \CA_{iL}(\shat) \vert^2
+ \vert \CA_{iR}(\shat) \vert^2 \ts,
\eea
where, for $\lambda = L,R$,
\bea\label{eq:zfactors}
\CA_{i\lambda}(\shat) &=& Q_i + {2 \zeta_{i\lambda} \ts \cot 2\thw \over
   {\sin 2\thw}} \ts \biggl({\shat \over {\shat - \CM^2_Z}}\biggr)\ts, \nn\\
\zeta_{i L} &=& T_{3i} - Q_i \sin^2\thw \ts, \nn\\
\zeta_{i R} &=& - Q_i \sin^2\thw \ts.
\eea
Here, $Q_i$ and $T_{3i} = \pm 1/2$ are the electric charge and left-handed
weak isospin of fermion $f_i$. Also, $\CM^2_{W,Z} = M^2_{W,Z} - i \rshat \ts
\Gamma_{W,Z}(\shat)$, where $\Gamma_{W,Z}(\shat)$ is the weak boson's
energy-dependent width.~\footnote{Note, for example, that $\Gamma_Z(\shat)$
includes a $t\ol t$ contribution when $\shat > 4m_t^2$.}

The $\tom$ decay rates to fermions with $N_f$ colors are given by
\be\label{eq:tomff}
\Gamma(\omega_T \ra \ol f_i f_i) = {N_f \ts \alpha^2 p\over
{3 \atro \shat}} \ts \left((\shat - m_i^2)
\ts B_i^0(\shat) + 6 m_i^2\ts \CR e(\CB_{iL}(\shat)
\CB_{iR}^*(\shat))\right) \ts, 
\ee
where
\bea\label{eq:bfactors}
&B_i^0(\shat) &= \vert \CB_{iL}(\shat) \vert^2
+ \vert \CB_{iR}(\shat) \vert^2 \ts, \nn \\ \nn\\
&\CB_{i\lambda}(\shat) &= \left[Q_i - {4 \zeta_{i\lambda} \sin^2\thw \over
    {\sin^2 2\thw}} \biggl({\shat \over {\shat - \CM_Z^2}}\biggr)\right] \ts
(Q_U + Q_D)
\ts.
\eea

\section*{3. Color--Singlet Technihadron Production Rates}

In this section we list the cross sections for the hadron collider
subprocesses $q \ol q \ra V_T \ra \tpi\tpi$, $G\tpi$, and $f \ol f$. The
$e^-e^+$ cross sections are obtained from the corresponding $q_i \ol q_i$
ones by multiplying the latter by 3. To include $\tro$--$\tom$ interference
in these cross sections, we use the mixed $\gamma$--$Z^0$--$\troz$--$\tom$
propagator matrix $\Delta_0$. It is the inverse of~\footnote{Note sign
changes in the off--diagonal elements relative to the propagator matrices in
Ref.~\cite{tcsm}.}
\be\label{eq:gzprop}
\Delta_0^{-1}(s) =\left(\ba{cccc}
s & 0 & s f_{\gamma\tro} & s f_{\gamma\tom} \\
0 & s - \CM^2_Z  & s f_{Z\tro} & s f_{Z\tom} \\
s f_{\gamma\tro}  & s f_{Z\tro}  & s - \CM^2_{\troz} & 0 \\
s f_{\gamma\tom}  & s f_{Z\tom}  & 0 & s - \CM^2_{\tom} 
\ea\right) \ts.
\ee
To guarantee a photon pole at $s=0$, we assume only kinetic mixing between
the gauge bosons and technivector mesons. In setting the off--diagonal
$\troz$--$\tom$ elements of this matrix equal zero, we are guided by the
smallness of this mixing in QCD and by the desire to keep the number of
adjustable parameters in the TCSM small. This mixing can always be added if
desired. The properly normalized $G V_T$ couplings are
\be\label{eq:fgv}
f_{GV_T} = 2\ts\xi\ts {\rm Tr}\biggl(Q_{G_V}Q^\dagger_{V_T}\biggr)
\ts;
\ee
in particular, $f_{\gamma\tro} = \xi$, $f_{\gamma\tom} = \xi \ts (Q_U +
Q_D)$, $f_{Z\tro} = \xi \ts \cot 2\thw$, and $f_{Z\tom} = - \xi \ts (Q_U +
Q_D) \tan\thw$, where $\xi = \sqrt{\alpha/\atro}$.
In the charged sector, the $W^\pm$--$\tropm$ matrix is the inverse of
\be\label{eq:wprop}
\Delta_{\pm}^{-1}(s) =\left(\ba{cc} s - \CM^2_W & s f_{W\tro} \\ s
  f_{W\tro} & s - \CM^2_{\tropm} \ea\right) \ts,
\ee
where $f_{W\tro} = \xi/(2\sin\thw)$.

The rates for production of any technipion pair, $\pi_A\pi_B = W_L W_L$,
$W_L\tpi$, and $\tpi\tpi$, in the isovector ($\tro$) channel
are:\footnote{For $W^+W^-$ and $W^\pm Z^0$ production, one should add the
standard model $t$ and $u$--channel amplitudes to the technicolor--modified
$s$--channel amplitude. This is a small correction for a narrow $\tro$ when
one effectively is always near the pole, as in a hadron collider. For a
lepton collider far below the $\tro$ pole, these contributions are
important. Similarly, the terms involving $\Delta_{\gamma\gamma}$,
$\Delta_{ZZ}$ and $\Delta_{\gamma Z}$ in Eq.~(\ref{eq:pippim}) are important
only far below the pole. They are relevant for LEP, but not the Tevatron.}
\bea\label{eq:pippim}
& &{d\hat\sigma(q_i \ol q_i \ra \pi^+_A\pi^-_B)
  \over{d\that}} \nn\\
&&\quad = {\pi \alpha^2 \CC^2_{AB} 
  (4\shat p^2 -(\that-\uhat)^2) \over{12 \shat^2}} \ts \sum_{\lambda=L,R}
 \biggl\vert Q_i \left(\Delta_{\gamma\gamma}(\shat)
 + f^{-1}_{\gamma \tro}\Delta_{\gamma\tro} \right) \\
&&\quad + {2\zeta_{i\lambda} \cot 2\thw \over{\sin2\thw}} \left(
  \Delta_{ZZ}(\shat) + f^{-1}_{Z \tro} \Delta_{Z \tro}(\shat)\right)
+ \left({Q_i \cos 2\thw + 2\zeta_{i\lambda}\over{\sin 2\thw}}\right)
  \Delta_{\gamma Z}\biggr\vert^2  \ts, \nn
\eea
and
\bea\label{eq:pippiz}
{d\hat\sigma(u_i \ol d_i \ra \pi^+_A\pi^0_B)
  \over{d\that}} &=& 
{\pi \alpha^2 \CC^2_{AB} 
  (4\shat p^2 -(\that-\uhat)^2) \over{96\sin^4\thw \shat^2}} \nn\\
&& \ts\ts\times  \biggl\vert \Delta_{WW}(\shat) + f^{-1}_{W\tro}
\Delta_{W\tro}(\shat)\biggr\vert^2 \ts,
\eea
where $p = [(\shat - (M_A+M_B)^2) (\shat - (M_A-M_B)^2)]^\half/2\rshat$ is the
$\shat$--dependent momentum of $\pi_{A,B}$. As usual, $\that = M^2_A -
\rshat(E_A - p\cos\theta)$, $\uhat = M^2_A - \rshat(E_A + p\cos\theta)$,
where $\theta$ is the c.m. production angle of $\pi_A$. The factor $4\shat
p^2 -(\that-\uhat)^2 = 4 \shat p^2 \sin^2\theta$.
Because the $\tro$--$\tom$ mixing parameter $\epsilon_{\rho\omega}$ is
expected to be very small, the rates for $q_i \ol q_i \ra \tom \ra \pi^+_A
\pi^-_B$ are ignored here.

The production of a transverse gauge boson plus technipion, $G\tpi$, is
dominated by the $\tro$, $\tom$ poles. However, in the limit $M_{\tro,\ts
\tom} \ra \infty$, it still proceeds via the usual axial vector anomaly
process, $G_1 \ra G_2 \tpi$~\cite{ehs}. This contribution interferes only
with the $V_{V_T G_2\tpi}$ term in Eq.~(\ref{eq:genamp}). It is small in a
hadron collider, but may be nonnegligible in an $e^+e^-$ collider (such as
LEP) operating well below the resonances. As discussed in
Ref.~\cite{Lane:2002wb}, we include the anomaly contribution by simply adding
it to the corresponding technivector amplitude. The cross section in the
neutral channel is given by
\bea\label{eq:gpineutral}
&& {d\hat\sigma(q_i \ol q_i \ra G \tpi)
\over{d\that}} \nn \\
&& \quad = {\pi \alpha^2 \over{24 \shat}} \Biggl\{
\left(\vert\CG^{VG\tpi}_{iL}(\shat)\vert^2 +
      \vert\CG^{VG\tpi}_{iR}(\shat)\vert^2\right) \ts
      \left(\that^2 + \uhat^2 -2M^2_G M^2_{\tpi}\right) 
\\
&& \qquad + \left(\vert\CG^{AG\tpi}_{iL}(\shat)\vert^2 +
             \vert\CG^{AG\tpi}_{iR}(\shat)\vert^2\right) \ts
\left(\that^2 + \uhat^2 -2M^2_G M^2_{\tpi} + 4\shat M^2_G\right)\Biggr\} \ts, \nn
\eea
where, for quark helicities $\lambda = L,R$,
\bea\label{eq:gfactors}
\CG^{VG\tpi} _{i\lambda}(\shat) &=& \sum_{V_T = \troz,\tom}
{V_{V_T G \tpi}\over{M_V}}  \biggl(Q_i\ts \Delta_{\gamma V_T}(\shat) +
{2 \zeta_{i\lambda} \over{\sin 2\thw}} \ts \Delta_{Z V_T}(\shat)\biggr) \nn\\
&+&\ts {e\Ntc \over{8\pi^2 F_T}} \biggl[ V_{\gamma G \tpi}
\biggl(Q_i \Delta_{\gamma\gamma} + {2\zeta_{i\lambda}\over{\sin 2\thw}}
\Delta_{Z\gamma}\biggr) 
+ V_{Z G \tpi}
\biggl(Q_i \Delta_{\gamma Z} + {2\zeta_{i\lambda}\over{\sin 2\thw}}
\Delta_{ZZ}\biggr) \biggr] \ts; \nn\\\nn\\
\CG^{AG\tpi} _{i\lambda}(\shat) &=& \sum_{V_T = \troz,\tom}
{A_{V_T G \tpi}\over{M_A}} \biggl(Q_i\ts \Delta_{\gamma V_T}(\shat) +
{2 \zeta_{i\lambda} \over{\sin 2\thw}} \ts \Delta_{Z V_T}(\shat)\biggr) \ts.
\eea
The anomaly factors in Eq.~(\ref{eq:gfactors}) are
\be\label{eq:anomaly}
V_{G_1G_2\tpi} = {\rm Tr}\left[Q^\dagger_{\tpi}
\left( \{Q_{G_1},Q^\dagger_{G_2}\}_L + \{Q_{G_1},Q^\dagger_{G_2}\}_R
\right)\right] \ts,
\ee
where $Q_{G_{R,L}} = Q_{G_V} \pm Q_{G_A}$ in Eq.~(\ref{eq:charges}). They are
listed in Table~2. Note that $\that^2 + \uhat^2 -2M^2_G M^2_{\tpi} = 2\shat
p^2 (1+\cos^2\theta)$. The $G\tpi$ cross section in the charged channel is
given by (in the approximation of a unit CKM matrix)
\bea\label{eq:gpicharged}
{d\hat\sigma(u_i \ol d_i \ra G \tpi) \over{d\that}} &=&
{\pi \alpha^2 \over{48 \sin^2\thw \ts \shat}} \ts \Biggr\{
{A^2_{\trop G\tpi}\over{M^2_A}} \left\vert\Delta_{W\tro}\right\vert^2
\biggl(\that^2 + \uhat^2 -2M^2_G M^2_{\tpi} + 4\shat M^2_G\biggr) \nn \\ \\
&+& \left\vert{e \Ntc \over{8\pi^2 F_T}} V_{W^+G\tpi} \Delta_{WW} +
{V_{\trop G\tpi}\over{M_V}} \Delta_{W\tro}\right\vert^2 
\biggl(\that^2 + \uhat^2 - 2M^2_G  M^2_{\tpi}\biggr)\Biggl\} \ts. \nn
\eea

{\begin{table}{
\begin{tabular}{|c|c|}
\hline
Process & $V_{G_1 G_2\tpi}$ 
\\
\hline\hline
$\gamma \ra \gamma \tpiz$& $2(Q_U + Q_D)\ts c_\chi$ \\
$\ts\ts\ts\quad \ra \gamma \tpipr$ & $2(Q_U^2 + Q_D^2)\ts c_{\chipr}$ \\ 
$\qquad \ra Z^0 \tpiz$ & $(Q_U + Q_D)\ts c_\chi\ts (1-4\sin^2\thw)/\sin
2\thw$ \\ 
$\ts\qquad \ra Z^0 \tpipr$ & $c_{\chipr}\ts [1-4(Q_U^2 + Q_D^2)\sin^2\thw]/
\sin 2\thw$ \\ 
$\ts\ts\ts\ts\qquad \ra W^\pm \tpimp$ & $(Q_U+Q_D)\ts c_\chi/(2\sin\thw)$ \\ 
\hline
$Z^0 \ra \gamma \tpiz$ & $(Q_U + Q_D)\ts c_\chi\ts (1-4\sin^2\thw)/\sin
2\thw$ \\ 
$\ts\ts\ts\quad \ra \gamma \tpipr$ & $c_{\chipr}\ts
[1-4(Q_U^2 + Q_D^2)\sin^2\thw]/\sin 2\thw$ \\
$\qquad \ra Z^0 \tpiz$ & $-(Q_U+Q_D)\ts c_\chi \cos 2\thw/\cos^2\thw$ \\
$\ts\qquad \ra Z^0 \tpipr$ & $2 c_{\chipr} [\cos 2\thw + 4(Q_U^2 + Q_D^2)
\sin^4\thw]/\sin^2 2\thw
 $ \\
$\ts\ts\ts\ts\qquad \ra W^\pm \tpimp$ & $-(Q_U + Q_D)\ts c_\chi
/(2\cos\thw)$\\ 
\hline
$W^\pm \ra \gamma \tpipm$ & $(Q_U + Q_D)\ts c_\chi/(2\sin\thw)$ \\ 
$\qquad\ts\ts\ts \ra Z^0 \tpipm$ & $-(Q_U+Q_D)\ts c_\chi/(2\cos\thw)$ \\  
$\ts\ts\ts\ts\ts\ts\qquad \ra W^\pm \tpiz$ & 0 \\ 
$\ts\ts\ts\ts\ts\ts\qquad \ra W^\pm \tpipr$ & $c_{\chipr}/(2\sin^2\thw)$ \\
\hline\hline
\end{tabular}}
\caption{Anomaly factors for $G_1 \ra G_2 \tpi$ for $G_i$ a
  transverse electroweak boson, $\gamma,Z^0,W^\pm$. Here, $c_\chi = \cos\chi$
  and $c_{\chipr} = \cos\chipr$.}
\end{table}}

The cross section for $q_i \ol q_i \ra f_j \ol f_j$ in the {\it
color--singlet} channel (with $m_{q_i} = 0$ and allowing $m_{j} \ne 0$ for
  $t \ol t$ production) is
\bea\label{eq:qqffrate}
{d\hat\sigma(q_i \ol q_i \ra \gamma ,\ts Z \ra \ol f_j f_j)
  \over{d\that}} &=& 
{N_f \pi \alpha^2\over{3\shat^2}} \biggl\{\left((\uhat-m_{j}^2)^2 +
  m_{j}^2\shat\right)
\ts \left(\vert\CD_{ijLL}\vert^2 + \vert\CD_{ijRR}\vert^2\right) \nn\\
&+& \left((\that-m_{j}^2)^2 +
  m_{j}^2\shat\right)\ts\left(\vert\CD_{ijLR}\vert^2 + 
  \vert\CD_{ijRL}\vert^2\right)\biggr\} \ts, 
\eea
\nin where
\bea\label{eq:dfactors}
\CD_{ij\lambda\lambda'}(\shat) &=& Q_i Q_j \ts
\Delta_{\gamma\gamma}(\shat)  + {4\over{\sin^2 2\thw}} \ts \zeta_{i \lambda}
\ts \zeta_{\j \lambda'} \ts \Delta_{ZZ}(\shat) \\
&& + {2\over{\sin 2\thw}} \ts \biggl(\zeta_{i \lambda} Q_j
\Delta_{Z\gamma}(\shat) + Q_i \zeta_{j \lambda'} \Delta_{\gamma
  Z}(\shat)\biggr)
\ts. \nn 
\eea
Finally, the rate for the subprocess $u_i \ol d_i \ra f_j \ol f'_j$ is
\be\label{eq:udffrate}
{d\hat\sigma(u_i \ol d_i \ra W^+ \ra  f_j \ol f'_j)
  \over{d\that}} = {N_f \pi \alpha^2\over{12\sin^4\thw \ts \shat^2}} \ts
(\uhat - m^2_j)(\uhat-m^{'2}_j) \ts \vert\Delta_{WW}(\shat)\vert^2 \ts.
\ee

\section*{4. The Color--Nonsinglet Sector of the TCSM}

The elementary particles of our TC2 model are the $\suone \otimes \sutwo$
gauge bosons $V_1^A$ and $V_2^A$ ($A=1,\dots,8$) and the technifermion
doublets $T_1 = (U_1,D_1) \in ({\bf 3},{\bf 1},{\bf \Ntc})$ and $T_2 =
(U_2,D_2) \in ({\bf 1},{\bf 3},{\bf \Ntc})$ of $\suone \otimes \sutwo \otimes
\sutc$. The gauge boson mass eigenstates are the $\suc$ gluons $g^A$ and the
coloron octet $V_8^A$:
\bea\label{eq:gauge}
g^A &=& {g_2 V_1^A + g_1 V_2^A \over{\sqrt{g_1^2 + g_2^2}}} \equiv \sin\thc
\ts V_1^A + \cos\thc \ts V_2^A \ts, \nn\\ \nn\\
V_8^A &=& {g_1 V_1^A - g_2 V_2^A \over{\sqrt{g_1^2 + g_2^2}}} \equiv \cos\thc
\ts V_1^A - \sin\thc \ts V_2^A \ts,
\eea
where $\alpha_1 = g_1^2/4\pi \gg \alpha_2 \simeq \alpha_C = g_C^2/4\pi$ and
$g_C = g_1 g_2/\sqrt{g_1^2 + g_2^2}$. We shall assume that, near 1~TeV,
$\suone$ interactions are nearly strong enough to cause top quark
condensation. In the Nambu--Jona-Lasinio approximation, this means
$\alpha_{V_8} \equiv 4\alpha_C/\sin^2 2\thc \simge \pi/2 C_2({\bf
3})\cos^4\thc$ where $C_2({\bf 3}) = 4/3$ (see Ref.~\cite{blt}). With
$\alpha_C \simeq 0.1$, this requires  $\tan\thc
\simle \sqrt{0.08}$. In terms of the $\suone$ and $\sutwo$ currents
$j_{1,2\mu}^A$, the fermion--gluon--coloron interactions are
\be\label{eq:currents}
g_1 \ts j_{1\mu}^A V_1^{A\mu} + g_2 \ts j_{2\mu}^A V_2^{A\mu} = 
g_C\left(\cot\thc \ts j_{1\mu}^A - \tan\thc \ts
  j_{2\mu}^A\right) V_8^{A\mu} + g_C \ts j_{c\mu}^A \ts g^{A\mu} \ts,
\ee
where $j_{c\mu}^A = j_{1\mu}^A + j_{2\mu}^A$ is the $\suc$ current.

We consider the couplings of both standard~\cite{tctwohill} and
flavor--universal~\cite{ccs} TC2 models. In standard TC2, top and bottom
quarks couple to $\suone$ and the four light quarks to $\sutwo$. The fermion
parts of the $SU(3)_{1,2}$ currents are
\bea\label{eq:standard}
j_{1\mu}^A &=& \half \ol T_1 \gamma_\mu \lambda_A T_1 + \sum_{a=t,b}
\half \ol q_a \gamma_\mu \lambda_A q_a \ts;\nn\\
j_{2\mu}^A &=& \half \ol T_2 \gamma_\mu \lambda_A T_2 + \sum_{a=u,d,c,s}
\half \ol q_a \gamma_\mu \lambda_A q_a \ts.
\eea
In this case, colorons decay strongly to top and bottom quarks and weakly to
the light quarks. In flavor--universal TC2 all quarks couple to $\suone$, not
$\sutwo$, so that colorons couple to all flavors with equal strength.

We assume that $\sutc$, not topcolor $\suone\otimes\uone$, interactions
dominate the formation of color--nonsinglet $\tro$ and $\tpi$ states. This is
an assumption of convenience; it may be inconsistent with the requirement
that $\uone$ is instrumental in driving the nondiagonal form of the
condensates in Eq.~(\ref{eq:Tconds}). On firmer ground, we assume that
techni--isospin is not badly broken by ETC interactions so that the $\troct$
are isosinglets. Then, a useful initial basis for the ground state $\troct$'s
that mix with $g$ and $V_8$ is:
\bea\label{eq:technirhos}
& &\vert \rho_{11}^{A} \ts \rangle = \left(\lambda_A/2\right)_{\alpha\beta}
\vert   U_{1\alpha} \ol U_{1\beta} + D_{1\alpha} \ol D_{1\beta} \ts
\rangle_{1^-} \nn\\ 
& &\vert \rho_{22}^{A} \ts \rangle = \left(\lambda_A/2\right)_{\alpha\beta}
\vert   U_{2\alpha} \ol U_{2\beta} + D_{2\alpha} \ol D_{2\beta} \ts
\rangle_{1^-} \nn\\ 
& &\vert \rho_{12}^{A} \ts \rangle =
  \left(\lambda_A/2\sqrt{2}\right)_{\alpha\beta} \vert U_{1\alpha} \ol
  U_{2\beta} + D_{1\alpha} \ol D_{2\beta} + (1\leftrightarrow 2)\ts
  \rangle_{1^-}
  \nn\\
& &\vert \rho_{12'}^{A} \ts \rangle =
  i\left(\lambda_A/2\sqrt{2}\right)_{\alpha\beta} \vert
  U_{1\alpha} \ol U_{2\beta} + D_{1\alpha} \ol D_{2\beta} - (1\leftrightarrow
  2)\ts \rangle_{1^-}
\eea
The first two of these states, $\troctaa$ and $\troctbb$, mix with $V_8$ and
$g$. The topcolor--breaking condensate in Eq.~(\ref{eq:Tconds}), $\langle \ol
T_{1L} T_{2R} \rangle \neq 0$, causes them to mix with $\troctab$ and
$\troctabp$ also. The $\troct$ mixing with $g$ and $V_8$ is purely kinetic
and is governed by the $U(2)$ matrices $W_{L,R}^{U,D}$ that diagonalize the
$\ol T_{Li} T_{Rj}$ condensates.~\footnote{The condensate in
Eq.~(\ref{eq:Tconds}) breaks $\suone \otimes \sutwo$ down to ordinary color,
$\suc$. The unitary aligning matrices $W^{U,D} = W_L^{U,D} \ts
(W_R^{U,D})^\dag$. See Ref.~\cite{vacalign} for a recent discussion of vacuum
alignment in technicolor, the physics that gives rise to $W_{L,R}^{U,D}$. It
is likely that all elements of these matrices are comparable in
magnitude. For economy in the default parameters, we shall take
$W_R^{U,D}=1$, so that $W^{U,D} = W_L^{U,D}$.} These
matrices appear, e.g., in the $\suone$ current as follows:\footnote{In
standard TC2, these currents are parity--violating if $W_L \neq W_R$. In
flavor--universal TC2, the quark currents are purely vectorial because the
$W_{L,R}$ are unitary. The same is always true of the $\suc$ currents.}
\be\label{eq:current} j_{1\mu}^A = \half \ts \sum_{T=U,D}\ts 
\ol T_{iL}\ts W_{Li1}^{T\dag} \ts \gamma_\mu \ts \lambda_A \ts W_{L1j}^T \ts
T_{jL}\ts  + \ts (L \ra R) \ts + \cdots \ts.
\ee

Even though technicolor may be the strongest of $T_{1,2}$'s interactions, the
topcolor $\suone\otimes\uone$ interactions are not weak perturbations; they
must be strong enough near 1~TeV to condense top quarks. Therefore, the
$\uone$ interactions must be custodial--isospin invariant in the technifermion
sector~\cite{cdt,tctwoklee}. On the other hand, the $\sutwo$ interactions of
$T_2$ are relatively weak. Thus, the approximate chiral symmetry of $T_1$ and
$T_2$ is
\be\label{eq:chisymm} \left[SU(2)_L \otimes SU(2)_R\right]_1 \otimes
\left[SU(6)_L \otimes SU(6)_R\right]_2 \ts.
\ee
Then technifermion condensation leads to a number of (pseudo)Goldstone boson
technipions. Although we assume that the technirhos are too light to decay
into a pair of technipions, they may decay into one of them plus a gluon. The
relevant technipions are an isosinglet $\suc$ octet $\ol T_2 T_2$ state and
the color--singlet $\tpipr$ discussed in Section~1. The color--octet state
is~\footnote{This state is essentially the same as the $\eta_T$ appearing,
e.g., in the one--family technicolor model~\cite{ehlq}. However, and this is
an important feature of TC2 models, it does not have a strong coupling to
$\ol tt$ and, therefore, does not appear as a resonance in $\ol tt$
production as described in Ref.~\cite{etat}.}
\be\label{eq:technipis}
\vert \tpioct^{A} \ts \rangle = \left(\lambda_A/2\right)_{\alpha\beta}
\vert U_{2\alpha} \ol U_{2\beta} + D_{2\alpha} \ol D_{2\beta} \ts
\rangle_{0^-} \ts.
\ee
In the absence of $\troct$ mixing, only $\troctbb$ decays into one of these
technipions plus a gluon. The strength of $\troctbb \ra g\tpipr$ is
controlled by $\cos\chi''$, where $\chi''$ is another mixing angle.

For the straw man model, $\tpioct$ is assumed to decay into either
fermion--antifermion pairs or two gluons, with the decay rates:
\bea\label{eq:tpidecay}
 \Gamma(\tpioct \ra \ol f f) &=& {1 \over {4\pi F^2_T}}
 \ts N_f \ts p_f \ts C^2_{8f} \ts m_f^2 \nn\\ \nn\\
 \Gamma(\tpioct \ra gg) &=& {1 \over {128 \pi^3 F^2_T}} 
 \ts \alpha_C^2 \ts C^2_{8g} \ts \Ntc^2 \ts M_{\tpioct}^3 \ts.
\eea
Here, $C_{8f}$ is an ETC-model dependent constant. In Ref.~\cite{tctwokl} the
colored technifermions $T_{1,2}$ do not couple to quarks and
leptons. Accordingly, we take the default value $C_{8f} =0$. To turn on
$\tpioct \ra \ol b b$, set $C_{8b} \simeq 1$. In TC2 models, the $\ol t t$
mode has $C_{8t} \simeq m_b/m_t$. The number of colors of fermion~$f$ is
$N_f$; its momentum is $p_f$; $\alpha_C$ is the QCD coupling evaluated at the
technipion mass; and $C^2_{8g}$ is an $\suc$ Clebsch of order one. We use the
color--singlet sector technipion decay constant, $F_T = F_\pi \sin\chi$. The
default values of these and other parameters are tabulated at the end of this
paper.

\section*{5. $\troct$ and $V_8$ Decay Rates}

In low--scale technicolor, color--triplet and octet technipion masses are
expected to be a few hundred GeV. As noted, we expect the $\troct$ to decay
into $\ol q q$ and $gg$ dijets and $g \tpi$. The energy--dependent
two--parton decay rates are (for $\rho_{ij}$ mass
$\sqrt{s}$)\footnote{Zerwekh and Rosenfeld have argued that hidden local
symmetry implies $\troct \ra gg$ is forbidden~\cite{ZR}. However, as
Chivukula, Grant and Simmons have shown, hidden local symmetry is
inappropriate for bound state $\troct$'s. The latter authors do emphasize
that there is some uncertainty in the strength of the $\troct \ra gg$
amplitude~\cite{cgs}. Nevertheless, we use the canonical decay rate from
Ref.~\cite{ehlq}.} 
\bea\label{eq:trhodecay}
\Gamma(\rho_{ij} \ra \ol q_a q_a) &=& {\alpha_C \over{6}} \left\vert \xi_g
\ts  \delta_{ij} +
 {g_a \ts \xi_{\rho_{ij}} \ts s \over{g_C (s - \CM^2_{V_8})}}
\right\vert^2 \ts \left(1 + {2m_a^2 \over{s}} \right) \ts
\left(s - 4m_a^2\right)^\half \ts, \nn\\ \nn\\ 
\Gamma(\rho_{ij} \ra gg) &=& {\alpha_C^2 \sqrt{s} \ts\delta_{ij} \over {4
    \atro}} \ts,
\eea
where $\CM^2_{V_8} = M^2_{V_8} - i \sqrt{s} \ts \Gamma_{V_8}(s)$, and
$\Gamma_{V_8}(s)$ is defined in Eq.~(\ref{eq:vdecay}) below. The couplings in
Eq.~(\ref{eq:trhodecay}) are obtained from
Eqs.~(\ref{eq:currents}--\ref{eq:current}); for $W_{Rij}^{U,D} =
\delta_{ij}$, they are:
\bea\label{eq:trhocouplings}
g_a &=& \left(\ba{l} g_C \cot\thc \ts\ts\ts {\rm for} \ts\ts\ts q_a = t,b \\
-g_C \tan\thc \ts\ts\ts {\rm for} \ts\ts\ts q_a = u,d,c,s \ea\right)
\qquad ({\rm standard \ts\ts TC2}) \nn\\\nn\\
g_a &=& g_C \cot\thc \hskip 1.64truein ({\rm flavor \ts\ts universal \ts\ts
  TC2})\nn\\ 
\atro &=& {g_{\rho_T}^2\over{4\pi}} = 2.91 \left({3\over{\Ntc}}\right) \nn\\
\xi_g &=& {g_C\over{g_{\tro}}} \nn\\
\xi_{\rho_{11}} &=& {2g_C\over{g_{\tro} \sin2\thc}}\ts
\left[\fourth\left(\left\vert W_{L11}^U \right\vert^2 + \left\vert W_{L11}^D
    \right\vert^2 + 2\right) - \sin^2\thc \right] \nn\\
\xi_{\rho_{22}} &=& {2g_C\over{g_{\tro} \sin2\thc}}\ts
\left[\fourth\left(\left\vert W_{L12}^U \right\vert^2 + \left\vert W_{L12}^D
    \right\vert^2 \right) - \sin^2\thc \right] \nn\\
\xi_{\rho_{12}} &=&  {g_C\over{\sqrt{2}g_{\tro} \sin2\thc}}\ts
{\rm Re}\left[W_{L11}^{U*} \ts W_{L12}^U  + W_{L11}^{D*} \ts
  W_{L12}^D\right] \nn\\ 
\xi_{\rho_{12'}} &=& {g_C\over{\sqrt{2}g_{\tro} \sin2\thc}}\ts
{\rm Im}\left[W_{L11}^{U*} \ts W_{L12}^U  + W_{L11}^{D*} \ts
  W_{L12}^D\right] \ts.
\eea
It is clear from these expressions how to generalize to $W_R^{U,D} \neq
1$. We have assumed that the $\troct \ra \tpi\tpi$ coupling $g_{\rho_T}$ is
the same for all ground--state technirhos, and the same that we have used for
color singlets; as usual, we use the naive large--$\Ntc$ value. Using
Eq.~(\ref{eq:unitarity}), the $\xi_{\troct}$ parameters become
\bea\label{eq:trhocoupb}
\xi_{\rho_{11}} &=& {2g_C\over{g_{\tro} \sin2\thc}}\ts
\left[\fourth\left(2+ \left(W_{L11}^U \right)^2 + \left(W_{L11}^D
    \right)^2 \right) - \sin^2\thc \right] \nn\\
\xi_{\rho_{22}} &=& {2g_C\over{g_{\tro} \sin2\thc}}\ts
\left[\fourth\left(2 - \left(W_{L11}^U \right)^2 - \left(W_{L11}^D
    \right)^2 \right) - \sin^2\thc \right] \\
\xi_{\rho_{12}} &=&  {g_C\over{\sqrt{2}g_{\tro} \sin2\thc}}\ts
\left[W_{L11}^{U} \ts \sqrt{1-(W_{L11}^U)^2} \ts \cos\phi_U  + 
      W_{L11}^{D} \ts \sqrt{1-(W_{L11}^D)^2} \ts \cos\phi_D\right] \nn\\ 
\xi_{\rho_{12'}} &=& {g_C\over{\sqrt{2}g_{\tro} \sin2\thc}}\ts
\left[W_{L11}^{U} \ts \sqrt{1-(W_{L11}^U)^2} \ts \sin\phi_U  + 
      W_{L11}^{D} \ts \sqrt{1-(W_{L11}^D)^2} \ts \sin\phi_D\right]\ts.\nn
\eea

Only the $\troctbb$ can decay to $g\tpi$; the rates are
\bea\label{eq:trotogtpi}
\Gamma(\troctbb \ra g\tpioct) &=& {5 \alpha_C p^3\over{9 M^2_8}}
\nn\\
\Gamma(\troctbb \ra g\tpipr) &=& {2 \alpha_C \cos^2\chi'' \ts p^3
\over{9 M^2_8}} \ts.
\eea
Here, $p = (s-M^2_{\tpi})/2\sqrt{s}$ is the gluon or $\tpi$ momentum and
$M_8$ is a mass parameter of order several hundred~GeV, analogous to the
parameter $M_V$ in Eq.~(\ref{eq:genamp}). So long as $\tpioct \ra gg$ is the
octet technipion's principal decay mode, it will be difficult, if not
impossible, to detect it in $\troctbb$ decays.

The $V_8$ colorons are expected to be considerably heavier than the
$\troct$~\cite{blt,ibehs}. Since they couple to
quarks, they can decay to dijets; see
Eqs~(\ref{eq:currents},\ref{eq:standard}). In both the standard and
flavor--universal models, colorons couple strongly to $\ol T_1 T_1$, but with
strength $g_C$ to $\ol T_2 T_2$. Since relatively light technipions are $\ol
T_2 T_2$ states, we expect $\Gamma(V_8 \ra \tpi\tpi) = \CO(\alpha_C)$ and
$\Gamma(V_8 \ra g\tpi) = \CO(\alpha_C^2)$. 
In this paper, we make the simplifying assumption
that such decays are irrelevant.~\footnote{A heavy $V_8$ may be able to decay
to technihadron pairs containing $\ol T_1 T_1$. We shall not attempt to
include these modes in the $V_8$'s energy--dependent width. This
approximation is good at Tevatron energies.} Then the $V_8$ decay rate is the
sum over open channels of
\be\label{eq:vdecay}
\Gamma(V_8 \ra \ol q_a q_a) = {\alpha_a \over{6}} \ts
\left(1 + {2m_a^2 \over{s}} \right) \ts \left(s - 4m_a^2\right)^\half \ts,
\ee
where $\alpha_a = g^2_a/4\pi$ was given in Eq.~(\ref{eq:trhocouplings}) for
the two types of TC2 models we consider.

\section*{6. Subprocess Cross Sections for Dijets}

The subprocess cross sections presented below assume massless initial--state
partons. They are averaged over initial spins and colors. These cross
sections require the propagator matrix describing mixing between the gluon,
$V_8$ and the $\troct$'s. We have assumed that the mixing between gauge
bosons and $\troct$ is purely kinetic, proportional to $s$. There are also
mixing terms $M^2_{ij,kl} = M^2_{kl,ij}$ between different $\troct$, induced
by $\ol T_1 T_2$ condensation. They are given by

\bea\label{eq:Mijkl}
M^2_{11,22} &=& {1\over{2}} \biggl[|W_{12}^U|^2 + |W_{12}^D|^2\biggr]
M^{\prime \ts 2}_8 =
{1\over{2}} \biggl[2 - \bigl(W_{L11}^U \bigr)^2 -
\bigl(W_{L11}^D\bigr)^2 \biggr] M^{\prime \ts 2}_8
\nn\\ 
M^2_{11,12} &=& {1\over{\sqrt{2}}} {\rm Re}\biggl[W_{11}^U \ts W_{12}^{U} +
W_{11}^D \ts W_{12}^{D} \biggr]M^{\prime \ts 2}_8 \nn\\
&=& {1\over{\sqrt{2}}} \biggl[W_{L11}^{U} \ts \sqrt{1-(W_{L11}^U)^2} \ts
  \cos\phi_U + W_{L11}^{D} \ts \sqrt{1-(W_{L11}^D)^2} \ts \cos\phi_D\biggr]
M^{\prime \ts 2}_8
\nn\\
M^2_{11,12'} &=& -{1\over{\sqrt{2}}} {\rm Im}\biggl[W_{11}^U \ts W_{12}^U +
W_{11}^D \ts W_{12}^D \biggr]M^{\prime \ts 2}_8 \nn\\
&=& -{1\over{\sqrt{2}}} \biggl[W_{L11}^{U} \ts \sqrt{1-(W_{L11}^U)^2} \ts
  \sin\phi_U + W_{L11}^{D} \ts \sqrt{1-(W_{L11}^D)^2} \ts \sin\phi_D\biggr]
M^{\prime \ts 2}_8 
\nn\\ \\
M^2_{22,12} &=& {1\over{\sqrt{2}}} {\rm Re}\biggl[W_{21}^U \ts W_{22}^U +
W_{21}^D \ts W_{22}^D \biggr]M^{\prime 2}_8 \nn\\
&=& -{1\over{\sqrt{2}}} \biggl[W_{L11}^{U} \ts \sqrt{1-(W_{L11}^U)^2} \ts
  \cos\phi_U + W_{L11}^{D} \ts \sqrt{1-(W_{L11}^D)^2} \ts \cos\phi_D\biggr]
M^{\prime \ts 2}_8 
\nn\\
M^2_{22,12'} &=& {1\over{\sqrt{2}}} {\rm Im}\biggl[W_{21}^U \ts W_{22}^U +
W_{21}^D \ts W_{22}^D \biggr]M^{\prime \ts 2}_8 \nn\\
&=& {1\over{\sqrt{2}}} \biggl[W_{L11}^{U} \ts \sqrt{1-(W_{L11}^U)^2} \ts
  \sin\phi_U + W_{L11}^{D} \ts \sqrt{1-(W_{L11}^D)^2} \ts \sin\phi_D\biggr]
M^{\prime \ts 2}_8
\nn\\
M^2_{12,12'} &=& -{1\over{2}} {\rm Im} \biggl[(W_{12}^U)^2 +
(W_{12}^D)^2\biggr] M^{\prime\ts 2}_8 \nn\\
&=& -{1\over{2}}\biggl[\bigl(1-(W_{L11}^U)^2\bigr)\sin{2\phi_U} +
    \bigl(1-(W_{L11}^D)^2\bigr)\sin{2\phi_D} \biggr] M^{\prime \ts 2}_8
\ts.\nn
\eea
Here, $M'_8$ is another technicolor scale mass parameter of order $M_8$. In
the second equalities, we assumed $W_R^{U,D}$ = 1 and used
Eq.~(\ref{eq:unitarity})). The
$g$--$V_8$--$\troctaa$--$\troctbb$--$\troctab$--$\troctabp$ propagator is the
inverse of the symmetric matrix
\be\label{eq:invprop}
D^{-1}(s) = \left(\ba{cccccc}
s & 0 & s \ts \xi_g & s \ts \xi_g & 0 & 0 \\ \\
0 & s - \CM^2_{V_8} & s \ts \xi_{\rho_{11}} & s \ts \xi_{\rho_{22}} & s
\ts \xi_{\rho_{12}} & s \ts \xi_{\rho_{12'}}\\ \\
s \ts \xi_g & s \ts \xi_{\rho_{11}} & s - \CM^2_{11}  & -M^2_{11,22} &
-M^2_{11,12} & -M^2_{11,12'}\\ \\
s \ts \xi_g & s \ts \xi_{\rho_{22}} & -M^2_{11,22} & s - \CM^2_{22}  &
-M^2_{22,12} & -M^2_{22,12'}\\ \\
0 & s \ts \xi_{\rho_{12}} & -M^2_{11,12} & -M^2_{22,12} &
s - \CM^2_{12} & -M^2_{12,12'}\\ \\
0 & s \ts \xi_{\rho_{12'}} & -M^2_{11,12'} & -M^2_{22,12'} &
 -M^2_{12,12'} & s - \CM^2_{12'} \\
\ea\right) \ts.
\ee
Here, $\CM^2_V = M^2_V - i \sqrt{s} \ts \Gamma_V(s)$ uses the
energy--dependent widths of the octet vector bosons. The following
combinations of propagators are used in the parton--parton cross sections
(here, $q_{a,b} = u,d,c,s$ or $t,b$):
\bea\label{eq:propterms}
\CD_{q_aq_b}(s) &=& D_{gg}(s) +
     \left({g_a + g_b\over{g_C}}\right) \ts D_{gV_8}(s) +
     {g_a g_b\over{g_C^2}} \ts D_{V_8V_8}(s) \nn\\
\Delta_{gg}(s) &=& s \ts D_{gg}(s) - 1 \nn\\
\CD_{q_ag}(s) &=& D_{gg}(s) + {g_a\over {g_C}} \ts D_{gV_8}(s) \nn\\
\Delta_{q_ag}(s) &=& s \ts \CD_{q_ag}(s) - 1 \ts.
\eea
The subprocess cross sections for $2\ra 2$ scattering of light quarks
$u,d,c,s,b$ and gluons are given by (here $q_a \neq q_b$):
\bea\label{eq:qgsigmaa} 
{d\hat \sigma(\ol q_a q_a \ra \ol q_b q_b) \over {d \that}} &=& 
{4 \pi \alpha_C^2 \over {9 \shat^2}} \left\vert \CD_{q_aq_b}(\shat)
\right\vert^2 (\uhat^2 + \that^2) \\
{d\hat \sigma(q_a q_b \ra q_a q_b) \over {d \that}} &=& 
{d\hat \sigma(\ol q_a \ol q_b \ra \ol q_a \ol q_b) \over {d \that}} = 
{4 \pi \alpha_C^2 \over {9 \shat^2}} \left\vert \CD_{q_aq_b}(\that)
\right\vert^2 (\uhat^2 + \shat^2) \\
{d\hat \sigma(\ol q_a q_a \ra \ol q_a q_a) \over {d\that}} &=&
{4 \pi \alpha_C^2 \over {9 \shat^2}} \ts
\Biggl\{\left\vert\CD_{q_aq_a} (\shat)\right\vert^2 \ts
\left(\uhat^2+\that^2\right) +
\left\vert\CD_{q_aq_a} (\that)\right\vert^2 \ts
\left(\uhat^2+\shat^2\right) \nn\\
&& \ts\ts - {2\over{3}} \ts {\rm Re}\left(\CD_{q_aq_a}(\shat) \ts
\CD^*_{q_aq_a}(\that)\right) \ts \uhat^2 \Biggr\} \\
{d\hat \sigma(q_a q_a \ra q_a q_a) \over {d\that}} &=& 
{d\hat \sigma(\ol q_a \ol q_a \ra \ol q_a \ol q_a) \over {d\that}} \nn\\
&=& {2 \pi \alpha_C^2 \over {9 \shat^2}} \ts
\Biggl\{\left\vert\CD_{q_aq_a} (\that)\right\vert^2 \ts
\left(\uhat^2+\shat^2\right) +
\left\vert\CD_{q_aq_a} (\uhat)\right\vert^2 \ts
\left(\that^2+\shat^2\right) \nn\\
&& \quad - {2\over{3}}\ts {\rm Re}\left(\CD_{q_aq_a}(\that)
\ts \CD^*_{q_aq_a}(\uhat)\right) \ts \shat^2 \Biggr\}  \\
{d\hat \sigma(gg \ra \ol q_a q_a) \over {d\that}} &=&
{9\over{32}} {d\hat \sigma(\ol q_a q_a \ra gg) \over {d\that}}  \nn \\
&=& {3 \pi \alpha_C^2 \over {8 \shat^2}}
\left\{{4\over{9}} \left({\uhat \over {\that}} + {\that \over {\uhat}}
\right) - \left({\that^2 + \uhat^2 \over {\shat^2}}\right) +
 {2\that\uhat \over{\shat^2}} \ts
\left\vert \Delta_{q_ag}(\shat)\right\vert^2 \right\}  \nn\\
\\
{d\hat \sigma(q_a g \ra q_a g) \over {d\that}} &=&
{\pi \alpha_C^2 \over {\shat^2}}
\left\{\left({\shat^2 + \uhat^2 \over {\that^2}}\right)
- {4\over{9}} \left({\uhat \over {\shat}} + {\shat \over {\uhat}}
\right) - {2\shat\uhat \over{\that^2}} \ts
\left\vert \Delta_{q_ag}(\that)\right\vert^2 \right\} \nn \\
{d\hat \sigma(gg \ra gg)\over{d\that}} &=&
{9\pi\alpha_c^2\over{32\shat^2}} \Biggl\{
  \ts\left\vert\Dggs\left({\that - \uhat\over{\shat}}\right) +
\Dggt\left[{\uhat^2\over{\shat^2}}\left({\shat-\uhat\over{\that}}\right) +
  {8\uhat\over{\shat}}\right] + {2\shat\over{\that}} \right\vert^2 \nn\\
&&\qquad\quad +\ts\ts\ts\left\vert\Dggs\left({\uhat-\that\over{\shat}}\right)+
\Dggu\left[{\that^2\over{\shat^2}}\left({\shat-\that\over{\uhat}}\right) +
  {8\that\over{\shat}}\right] + {2\shat\over{\uhat}} \ts \right\vert^2 \nn\\
&&\qquad\quad + \ts\ts {\rm Re}
\Biggl[\ts \left(\Dggss\left({\that-\uhat\over{\shat}}\right) +
\Dggts\left[{\uhat^2\over{\shat^2}}\left({\shat-\uhat\over{\that}}\right)
 + {8\uhat\over{\shat}}\right] + {2\shat\over{\that}} \right) \nn\\
&&\qquad\quad\qquad \times 
\left(\Dggs\left({\uhat-\that\over{\shat}}
\right) +
\Dggu\left[{\that^2\over{\shat^2}}\left({\shat-\that\over{\uhat}}\right)
 + {8\that\over{\shat}}\right] + {2\shat\over{\uhat}} \right)\Biggr] \nn\\
&&\qquad\quad+\ts\ts \left\vert \Dggs\left({\that-\uhat\over{\shat}}\right) +
  \Dggt\left({\uhat^2-\shat^2 \over{\shat^2}}\right) \right\vert^2 \nn\\
&&\qquad\quad+\ts\ts \left\vert \Dggs\left({\uhat-\that\over{\shat}}\right) +
  \Dggu\left({\that^2-\shat^2 \over{\shat^2}}\right) \right\vert^2 \nn\\
&&\qquad\quad +\ts\ts {\rm Re} \Biggl[
\ts\left(\Dggss\left({\that-\uhat\over{\shat}}\right) +
  \Dggts\left({\uhat^2-\shat^2 \over{\shat^2}}\right)\right) \nn\\
&&\qquad\quad\qquad \times
  \left(\Dggs\left({\uhat-\that\over{\shat}}\right) +
  \Dggu\left({\that^2-\shat^2 \over{\shat^2}}\right)\right)\Biggr] \nn\\
&&\qquad\quad + \left({\uhat^4 + \that^4\over{\shat^4}}\right)
\Biggl[\ts\left\vert \Dggt
  \left({\shat-\uhat\over{\that}}\right) +
  {2\shat\over{\that}} \right\vert^2
      + \left\vert \Dggu \left({\shat-\that\over{\uhat}}\right) +
  {2\shat\over{\uhat}} \right\vert^2 \nn\\
&&\qquad\quad + \ts\ts
{\rm Re}\left[\left(\Dggts\left({\shat-\uhat\over{\that}}\right) +
    {2\shat\over{\that}} \right)
      \left(\Dggu\left({\shat-\that\over{\uhat}}\right) +
    {2\shat\over{\uhat}} \right) \right]\Biggr]  \nn\\
&&\qquad\quad+\ts\ts 4 \left({\that\uhat\over{\shat^2}}\right)^2
  \left[\ts \left\vert\Dggt\right\vert^2 +
    \left\vert\Dggu\right\vert^2  + {\rm Re} \left(\Dggts \Dggu\right)
\right] \Biggr\}
    \ts. 
\eea
For massless partons, $\that = -\half \shat(1-\cos\theta)$ and $\uhat =
-\half \shat(1+\cos\theta)$ where $\theta$ is the cm scattering angle.

The cross sections for top quark production are:
\bea\label{eq:topsigma}
{d\hat \sigma(\ol q_a q_a \ra \ol t t) \over {d \that}} &=& 
{4 \pi \alpha_C^2 \over {9 \shat^4}} \ts \left\vert \shat\CD_{q_at}(\shat)
\right\vert^2 \left[\left(\uhat -m_t^2\right)^2 + \left(\that -
    m_t^2\right)^2 + 2m_t^2 \shat\right] \\
{d\hat \sigma(gg \ra \ol tt) \over {d\that}} &=&
{3\pi \alpha_C^2 \over {8 \shat^2}}
\Biggl\{{4\over{9}}
 \Biggr[{(\that-m_t^2)(\uhat-m_t^2)-2m_t^2(\that+m_t^2)\over
  {(\that-m_t^2)^2}} \nn\\
&&\qquad\quad +\ts\ts {(\that-m_t^2)(\uhat-m_t^2)-2m_t^2(\uhat+m_t^2)\over
  {(\uhat-m_t^2)^2}}\Biggr] \nn\\
&&\quad + \ts\ts {(\that-m_t^2)\left(\uhat-m_t^2\right) +
m_t^2\left(\that-\uhat\right)\ts {\rm Re}\left(\Delta_{tg}(\shat) - 1\right)
\over{\shat(\that-m_t^2)}} \nn\\
&&\quad +\ts\ts {(\that-m_t^2)(\uhat-m_t^2) +
m_t^2(\uhat-\that)\ts {\rm Re}\left(\Delta_{tg}(\shat) - 1\right)
\over{\shat(\uhat-m_t^2)}} \nn\\
&&\quad +\ts\ts {2\over{\shat^2}} (\that-m_t^2)(\uhat-m_t^2) \ts
\left[\left\vert \Delta_{tg}(\shat) \right\vert^2 +1\right] 
-\ts\ts {m_t^2\left(\shat-4m_t^2\right)\over{9\left(\that-m_t^2\right)
\left(\uhat-m_t^2\right)}} \Biggr\} \ts. \nn\\
\eea
Here, $\that = m_t^2 - \half \shat(1-\beta_t\cos\theta)$ where $\beta_t =
\sqrt{1-4m_t^2/\shat}$.

\section*{7. Hadron Collider Phenomenology}

In this section, we concentrate on the phenomenology of the color--octet
sector of the TCSM. The color--singlet sector has been studied in
Refs.~\cite{tcsm,Lane:2002wb,Mrenna:1999ks,l3,delphi,opal,CDFWtpi,CDFgtpi,ATLAS}. 
Numerical results are presented
for the Tevatron collider, since new strong interactions may
be visible there if they are related to electroweak symmetry breaking and the
top quark's mass. Our results are readily generalizable to the LHC.

The impact of the coloron $V_8$ on high-$p_T$ scattering has been explored in
the past~\cite{ibehs,cthsjp,kltop} and, by itself, it is easy to
understand. The improvement in the present approach is to include the full
propagator structure, with an energy-dependent width $\Gamma_{V_8}(s)$
for the coloron. The simplification of treating coloron-exchange as a contact
term is appropriate when the coloron is very heavy and
$\sqrt{s}\Gamma_{V_8}(s) \ll M^2_{V_8}$.

Let us first ignore the subleading effects of kinetic mixing with the
color--octet technirhos. Then, in fermion scattering processes, the coloron
appears as a modification to the gluon propagator which may depend on the
fermion generations. For standard TC2, there is an enhanced coloron coupling
to $b$ and $t$ quarks, which changes the gluon propagator connecting pairs of
fermions to
\be
{\cal D}_{qt}(s) = {1\over s}\left[1-{s\over s-\Mv^2 -
    i\sqrt{s}\Gamma_{V_8}(s)}\right]\ts.
\ee
In the limit $\Mv\gg m_t$, the energy dependent width scales as $\sqrt{s}$.
For the ``NJL--inspired'' value of the topcolor $SU(3)$ mixing angle,
$\tan\thc=\sqrt{.08}$, the coloron width is roughly $\sqrt{s}/3$. Below
resonance, the coloron interferes constructively, and can thus manifest
itself as an energy--dependent enhancement of the $b$ or $t$ quark cross
sections. The invariant mass distribution of top quark pairs $\CM_{\ol tt}$
was measured by both CDF and D\O~\cite{cdftop,d0top}, and it constrains the
values of $\Mv$ and $\tan\thc$ (the latter dependence arises indirectly
through the coloron width). Roughly, these measurements are consistent with a
coloron mass near 1~TeV, with little restriction on the value of
$\tan\thc$.~\footnote{This and the following statement for the
universal coloron are based on basic fits to the public Run~I data, but
a more systematic analysis is needed.
There are no published results on
limits for wide resonances in the $\ol tt$ system.}  From the $\CM_{\ol bb}$
distribution, CDF~\cite{Hoffman:mk} excludes standard TC2 colorons in the
mass range $280-670$ GeV for $\Gamma_{V_8}/M_{V_8}=0.3$. The limit worsens as
the width is increased, essentially vanishing for $\Gamma_{V_8}/M_{V_8}>0.7$.

For flavor--universal TC2, there is an enhanced coloron coupling to
all flavors of quarks: 
\be
{\cal D}_{qq'}(s) = {1\over s}\left[1+{1\over \tan^2\thc}{s\over s-\Mv^2
    - i\sqrt{s}\Gamma_{V_8}(s)}\right].
\label{eq:dqtu}
\ee
In contrast to standard TC2, the coloron interferes destructively below
resonance, almost cancelling the simple gluon exchange at
$\sqrt{s}=\Mv\sin\thc$. Furthermore, the on--mass--shell coloron width is
comparable to its mass for our default choice of parameters. We shall see
below that consistency with the CDF and D$\O$ data tends to require a
universal coloron mass of several TeV, in agreement with the D\O\ limit
resulting from dijet measurements~\cite{ibehs}. Despite the fact that the
interference occurs for all flavors of quarks in flavor--universal TC2, the
$\CM_{\ol tt}$ distribution itself ultimately may be the most constraining,
because of the relative importance of $\ol qq$ annihilation to other partonic
processes. In dijet cross sections, many different subprocesses contribute,
some with $s$--channel gluon exchanges (which are destructive below
resonance) and some with $t$ and $u$--channel exchanges (which are
constructive).

While the coloron phenomenology is straightforward, the effects of the
color-octet technirhos are less transparent. To simplify the discussion, we
consider first the case of a coloron and a single technirho mixing with the
gluon and coloron. The inverse propagator matrix describing the
gluon--$V_8$--$\troct$ mixing is
\be
\CD_0^{-1}(s) =\left(\ba{ccc}
s & 0 & s \xi \\
0 & s - \CM^2_{V_8}  & s \xi' \\
s \xi & s \xi'  & s - \CM^2_{\rho}  
\ea\right) \ts,
\ee
where $\xi = \xi_g$ and $\xi' = \xi_{\rho_{ij}}$ set the
strength of the kinetic mixing.
The cross sections depend on the propagators (see Eq.~\ref{eq:propterms}):
\bea
D_{gg} & = & {(s- \CM^2_{V_8} )(s - \CM^2_{\rho}) - s^2 \xi'^2 \over 
{\rm Det}\CD_0^{-1}} = {1\over s}+  {s(s- \CM^2_{V_8} )\xi^2 \over 
{\rm Det}\CD_0^{-1}}\\
D_{gV_8} & = & { s^2 \xi\xi' \over {\rm Det}\CD_0^{-1}} \\
D_{V_8V_8} & = & {s(s - \CM^2_{\rho}) - s^2 \xi^2 \over 
{\rm Det}\CD_0^{-1}} = {1\over (s- \CM^2_{V_8} )} +
{s^3\xi'^2 \over (s- \CM^2_{V_8}){\rm Det}\CD_0^{-1}}\\
{\rm Det}\CD_0^{-1} & = & s(s- \CM^2_{V_8} )(s - \CM^2_{\rho})-s^2(s\xi'^2
+ (s-\CM^2_{V_8})\xi^2)\ts.
\eea
When written in this form, the gluon and coloron components are
approximately separated from the technirho's.

The $\ol tt$ cross section involves ${\cal D}_{qt}$,
which depends on the specific model. For standard TC2, it has the form
\bea
{\cal D}_{qt}= {1\over s}-{1\over s-\CM^2_{V_8}} + \hbox{\hskip 5cm}\nonumber
\\
{1 \over {\rm Det}\CD_0^{-1} } \left[ s(s-\CM^2_{V_8})\xi^2
+ s^2\xi\xi'(-\tan\thc+{1\over \tan\thc})
-{s^3\xi'^2\over s-\CM^2_{V_8}}\right]\ts.
\eea
The technirho phenomenology is determined by the last term.
Taking $\rho=\rho_{22}$, e.g., this can be written as $\xi^2 s^2$ times
the dimensionless quantity
\be
{\cal N}=r-\tan^2\thc/r\left( {1-W^2 \over 2\sin^2\thc}-1\right)^2
 +(1-\tan^2\thc)\left( {1-W^2\over 2\sin^2\thc}-1\right)\ts,
\ee
with $r=1-\CM^2_{V_8}/s$ and we put $W = W^U_{11} = W^D_{11}$.

For small coloron--technirho mixing, $\xi\gg |\xi'|$, the technirho
contribution to ${\cal D}_{qt}$ reduces to $\xi^2/(s-\CM^2_\rho)$, yielding
the behavior of a narrow resonance. For comparable $\xi$ and $\xi'$, the
maximum of ${\cal D}_{qt}$ near $s=M_\rho^2$ is no longer just inversely
proportional to $\Gamma_\rho$, but also depends on
$\xi'^2\Gamma_{V_8}/r^2$. The effective technirho width is larger and the
cross section is smaller, depending on the mixing and coloron properties.
Furthermore, certain relations may exist among the parameters in ${\cal N}$
that weaken the resonant behavior. For example, the numerator vanishes for
$r=\xi'/\xi\tan\thc$ or $r=-\xi'/\xi\cot\thc$. We generally expect that
technirhos are lighter than the coloron, so only the latter solution is
relevant. This relation is difficult to satisfy unless $W$ is small. As an
example, for $W=0$, $r=(1/.4)^2$, and $\tan\thc=0.295$, we have ${\cal
N}=0$. Similar remarks hold in the case of the flavor--universal coloron,
where the simplified propagator has the form
\bea 
{\cal D}_{qt}= {1\over s}+{\cot^2\thc\over s-\CM^2_{V_8}} + 
\hbox{\hskip 5cm}\nonumber \\
{1\over {\rm Det}\CD_0^{-1}}\left[s(s-\CM^2_{V_8})\xi^2
+ 2 s^2\xi\xi'\cot\thc
+{s^3\xi'^2\cot^2\thc\over s-\CM^2_{V_8}}\right] \ts.
\eea
In particular, the same condition, $r=-\xi'/\xi \cot\thc$, makes the
numerator factor corresponding to $\CN$ vanish.

While this one--technirho example illustrates the propagator structure, the
full TC2 model's additional $\troct$'s and their mixing can lead to
substantially more complicated spectra. In addition, the $\rho_{22}$ can have
significant $g\tpipr$ and $g\tpioct$ decay rates that decrease the cross
section on resonance. The resultant phenomenology can be quite rich. Here, we
focus on a several benchmark models that demonstrate some interesting
features. For further background, we note that CDF placed limits on the
additional cross section from {\it narrow} resonances ($\Gamma/M<0.1$) in the
$\CM_{\ol bb}$~\cite{Hoffman:mk} and $\CM_{\ol tt}$~\cite{cdftop}
distributions. For invariant masses less than about 550 GeV, $\CM_{\ol bb}$
is more sensitive, while $\CM_{\ol tt}$ is more sensitive above. Neither of
these Run~I limits excludes a single, isolated technirho in either the $\ol
bb$ or $\ol tt$ channels. It is not a trivial matter to apply these limits to
the cases when several technirhos have comparable mass or when the coloron is
relevant.

\begin{figure}[!ht]
  \begin{center}
    \psfig{file=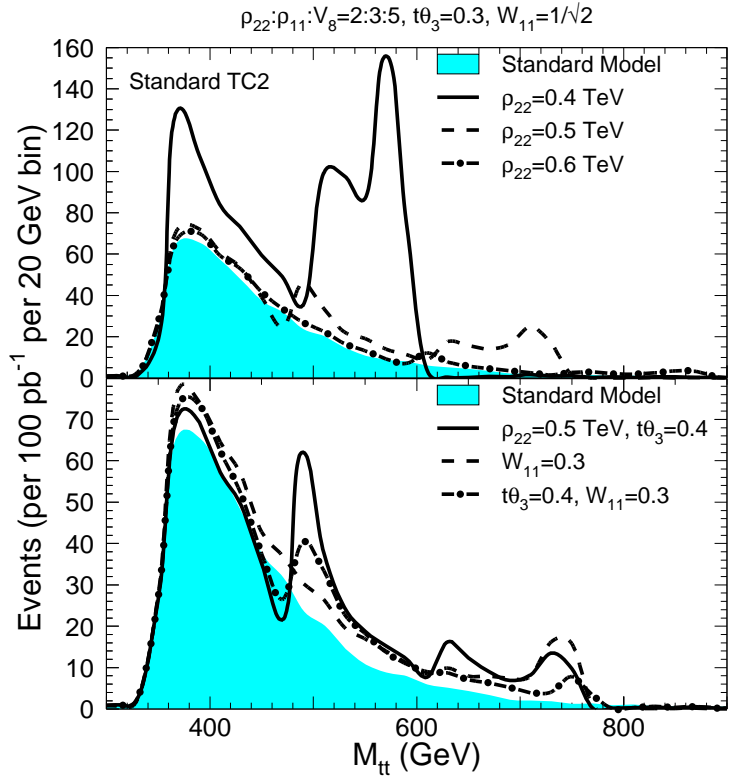,width=11cm}
    \caption{Invariant mass distributions of top quark pairs $\CM_{\ol tt}$
      compared to the standard model distribution for several choices of
      parameters in standard TC2.}
    \label{fig:bm1}
  \end{center}
\end{figure}

\leftline{{\bf Benchmark I:  Standard TC2---$\CM_{\ol t t}$ with various
    $M_{\troct}$}}

This benchmark studies effects from Standard TC2 on the invariant mass
distribution of $\ol tt$ pairs produced at the Tevatron in Run~I ($\sqrt{s} =
1.8\,\tev$ and $\int\CL dt = 100\,\ipb$). The curves were generated using
{\sc Pythia}, but only at the parton level, and with no efficiency or
resolution effects. The masses of the $\rho_{22}$, $\rho_{11}$ and $V_8$
were fixed in the ratio 2:3:5, with the $\rho_{12}$ and $\rho_{21}$
degenerate in mass at the average of $M_{\rho_{11}}$ and
$M_{\rho_{22}}$.~\footnote{This pattern of technirho masses is assumed
throughout the rest of this analysis, and is motivated by a valence
technifermion approximation.} We set the mass $M_{\tpioct}=300\,\gev$.
Initially, we choose $\tan\thc=0.3$ and $W =1/\sqrt{2}$; all other parameters
are given by their default values in Table~3. For normalization, the standard
model distribution is shown as the filled histogram in Fig.~1. As
$M_{\rho_{22}}$ (and hence all other masses from the fixed ratio) is
increased from 0.4~TeV (solid line, upper panel) to 0.5~TeV (dashes) to
0.6~TeV (dash--dot), narrow peaks occur with roughly the same relative
enhancement over the standard model distribution: the value of ${\cal N} =
-3.33$ is the same at $\sqrt{s}=M_{\rho_{22}}$ in each case. The absolute
size of the peaks decreases with increased mass because of the decreased
partonic luminosity. We also studied the dependence of these results on $W$
and $\tan\thc$ for a fixed value of $\rho_{22}=0.5\,\tev$ (lower panel).
First $\tan\thc$ is increased (solid) with all other parameters at the
baseline value, second $W$ is decreased (dashes) with all other parameters at
the baseline value, and third both $\tan\thc$ and $W$ are varied
simultaneously (dash--dot) with all other parameters fixed. In these last
three cases, the value of ${\cal N}$ at $\sqrt{s}=M_{\rho_{22}}$ varies from
$-4.55$ to $-0.80$ to $-3.16$, which is reflected in the relative height of
the peaks. The latter should be compared to the baseline value of ${\cal
N}=-3.33$. Note that none of the individual resonances is excludable with
Run~I data; a full shape analysis may be sensitive to the first set of
parameters (upper panel, solid line). A promising model--line for study by
the experiments is to vary $M_{\rho_{22}}$ with the mass ratio fixed and for
specific values of $W$ and $\tan\thc$. It would also be interesting to vary
the parameter $M'_8$, which induces hard mixing between the $\troct$'s in
Eq.~(\ref{eq:Mijkl}). As apparent from the plots, the technirho pole masses
are shifted from the value of the input mass parameter because of the
off--diagonal terms there.

One can also consider the $\ol bb$ final state to complement the $\ol tt$
distributions. While there are several sources of $\ol bb$ production, the
largest effect is expected to occur in the direct production mechanism. This
can be isolated by requiring two, high--$p_T$, $b$--tagged jets that are
balanced azimuthally ($\Delta\phi\sim\pi$). While such a search would give
sensitivity to $\troct$ states below the $\ol tt$ threshold, it would not
greatly improve the search for higher mass states.

\begin{figure}[!ht]
  \begin{center}
    \psfig{file=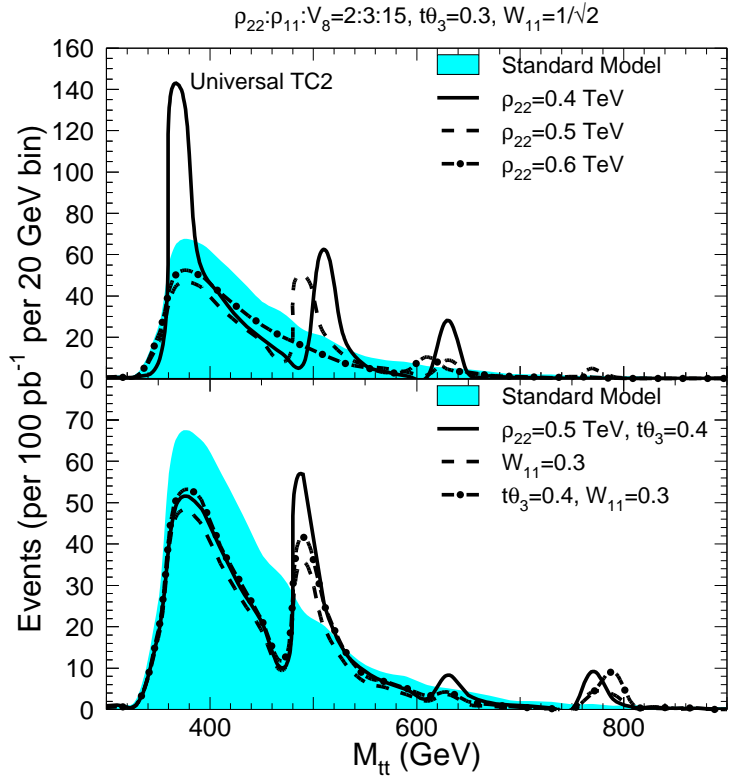,width=11cm}
    \caption{Invariant mass distributions of top quark pairs $\CM_{\ol tt}$
      compared to the standard model distribution for several parameter
      choices in flavor--universal TC2.} 
    \label{fig:bm2}
  \end{center}
\end{figure}

\leftline{{\bf Benchmark II:  Flavor--Universal TC2---$\CM_{\ol t t}$ with
    various $M_{\troct}$}}

This benchmark studies the impact of flavor--universal TC2. The parameter
choices are identical to the case of Benchmark~I for standard TC2, except
that the masses of the $\rho_{22}$, $\rho_{11}$ and $V_8$ fixed in the ratio
2:3:15, because of the coloron's stronger coupling in these models. The
results are displayed in Fig.~2. The $\CM_{\ol tt}$ distribution is distorted
both by the presence of narrow $\troct$'s and the destructive interference
between the gluon and coloron---even though the flavor--universal coloron has
a mass of 3--4.5~TeV! As in the first benchmark, none of the individual
technirho resonances can be excluded with Run~I data. The case
$M_{\rho_{22}}=0.4\,\tev$ corresponds to the $\Mv\tan\thc = 0.84$~TeV limit
set from the D\O\ dijet data by Bertram and Simmons~\cite{ibehs} without
including the $\troct$'s.

\begin{figure}[!ht]
  \begin{center}
    \psfig{file=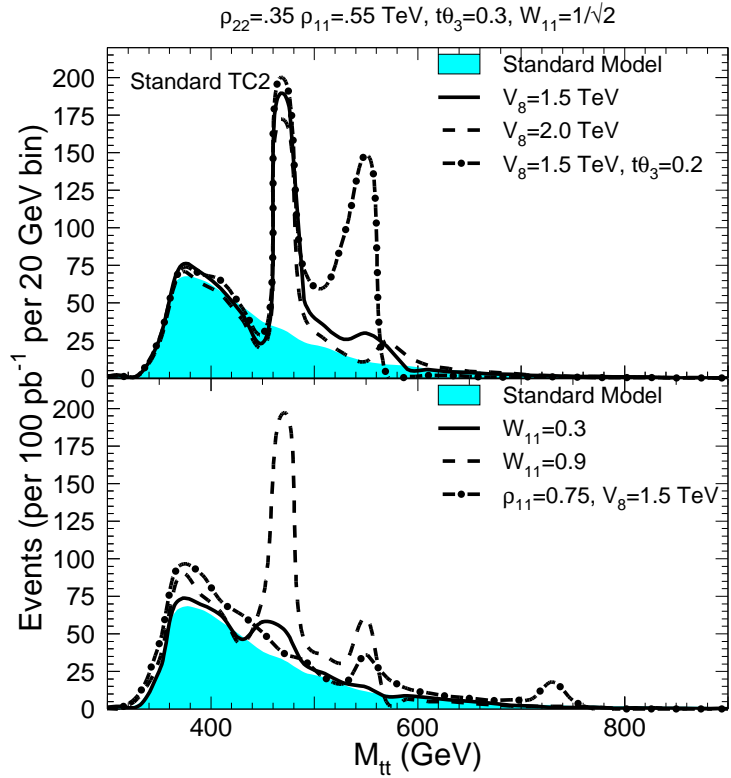,width=11cm}
    \caption{Similar to Fig.~\ref{fig:bm1}, but showing the
variation with coloron mass.}
    \label{fig:bm3}
  \end{center}
\end{figure}

\leftline{{\bf Benchmark III:  Standard TC2---$\CM_{\ol t t}$ with various
    $\Mv$}}

This benchmark demonstrates the variation with the coloron mass in standard
TC2. The technirhos are fixed at the mass values of
$M_{\rho_{22}}=.35\,\tev$, $M_{\rho_{11}}=.55\,\tev$, and
$M_{\rho_{12,21}}=.45\,\tev$. This is specifically chosen so that the
$\rho_{22}$ is below the $\ol tt$ threshold; the resultant peak in $\CM_{\ol
bb}$ (not shown here) is consistent with the Run~I limits discussed earlier
\cite{Hoffman:mk}. The results are shown in Fig.~\ref{fig:bm3}. For these
choices of parameters, narrow resonances appear near $M_{\rho_{11}}$ and
$M_{\rho_{21}}$. The appearance of two significant peaks for a lower value of
$\tan\thc=0.2$ (upper panel, dash--dot) may be excludable by a
multi--resonance analysis, but is presently allowed by the Run~I data. A
promising model--line for this benchmark is to vary the technirho mixing
through $W$ and $\tan\thc$ with all other masses fixed. For extreme values of
the coloron mass, this will become similar to Benchmarks~V and~VI described
later.

\begin{figure}[!ht]
  \begin{center}
    \psfig{file=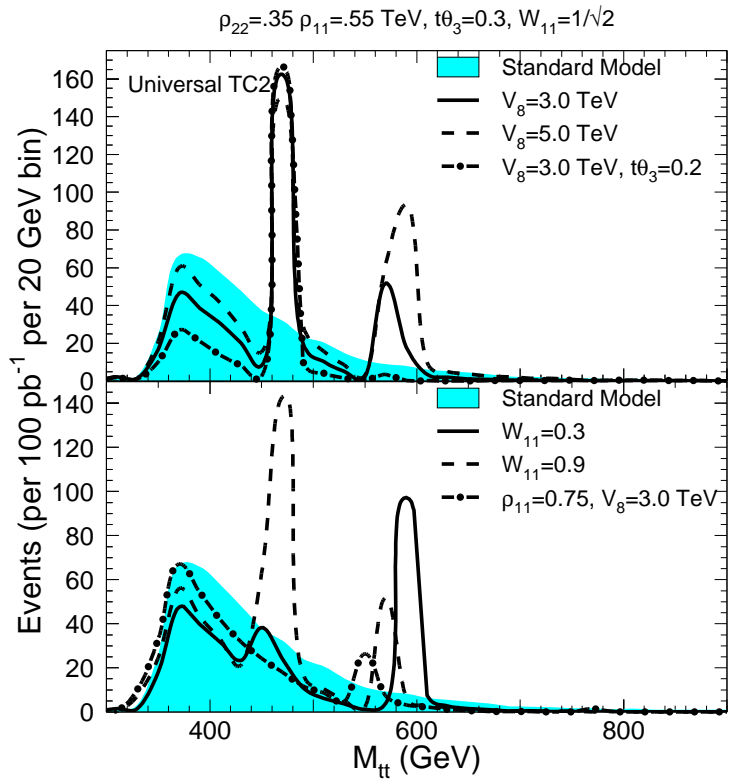,width=11cm}
    \caption{Similar to Fig.~\ref{fig:bm2}, but showing the
variation with coloron mass.}
    \label{fig:bm4}
  \end{center}
\end{figure}

\leftline{{\bf Benchmark IV:  Flavor--Universal TC2---$\CM_{\ol t t}$ with
    various $\Mv$}}

The parameter choices here are similar to that of Benchmark~III for standard
TC2, except that the colorons are heavier. The results are shown in
Fig.~\ref{fig:bm4}. As in Benchmark~III, there are narrow resonances, but
destructive rather than constructive gluon--coloron interference. The
standard model invariant mass distribution is greatly distorted, except for
the case of a heavier $\rho_{11}$ (lower panel, dash--dot). The coloron mass
$M_{V_8} = 3\,\tev$ in this case is consistent with the limit obtained by
Bertram and Simmons.

\begin{figure}[!ht]
  \begin{center}
    \psfig{file=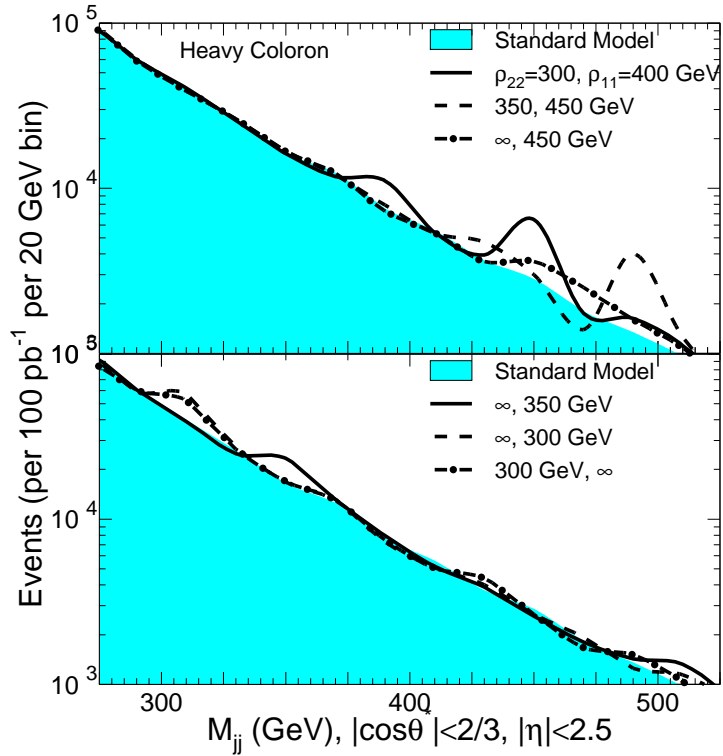,width=11cm}
    \caption{Dijet Invariant mass distributions 
compared to the standard model distribution for flavor
universal TC2 with a heavy coloron, but light technirhos.}
    \label{fig:cdf}
  \end{center}
\end{figure}

\leftline{{\bf Benchmark V---Universal TC2 jets with large $\Mv$}}

Our last two benchmarks illustrate the effect of flavor--universal TC2 on
light--parton dijet mass distributions. For the first, the coloron is taken
heavy, and we focus on the lighter technirho resonances. This corresponds to
the CDF search in Run~I for a single (pre--TC2) color octet
technirho~\cite{CDFjets}. Dijets in this study were required to have
$|\cos\theta| < 2/3$ and $|\eta| < 2.5$ for the cms scattering angle and
rapidity. The search excluded a $\troct$ resonance in the mass range of
$260-480$ GeV. The standard model prediction is shown in Fig.~\ref{fig:cdf}
by the solid histogram. The first two curves are for technirho masses of
$M_{\rho_{22,11}}=300,400$ GeV (solid) and $M_{\rho_{22,11}}=350,450$ GeV
(dashes). As before, $M_{\rho_{12,21}}$ is taken as the average of
these. Significant features appear in the invariant mass distribution, but
shifted from the input $M_{\rho_{ij}}$ because of mixing effects in the
$6\times 6$ propagator matrix. The 300~GeV $\rho_{22}$ sits on too much
background to be visible, i.e., in this model it could not be excluded by the
CDF search in Run~I. The last four curves treat the case of only one light
technirho and, as expected, they exhibit resonant structure quite near the
input mass. A promising model--line is to start with nearly equal values for
$M_{\rho_{22}}$ and $M_{\rho_{11}}$, and slowly increase $M_{\rho_{11}}$
until the single technirho limit is reached. Also, the sensitivity to mass
splitting and the mixing parameter $M'_8$ should be considered.

\begin{figure}[!ht]
  \begin{center}
    \psfig{file=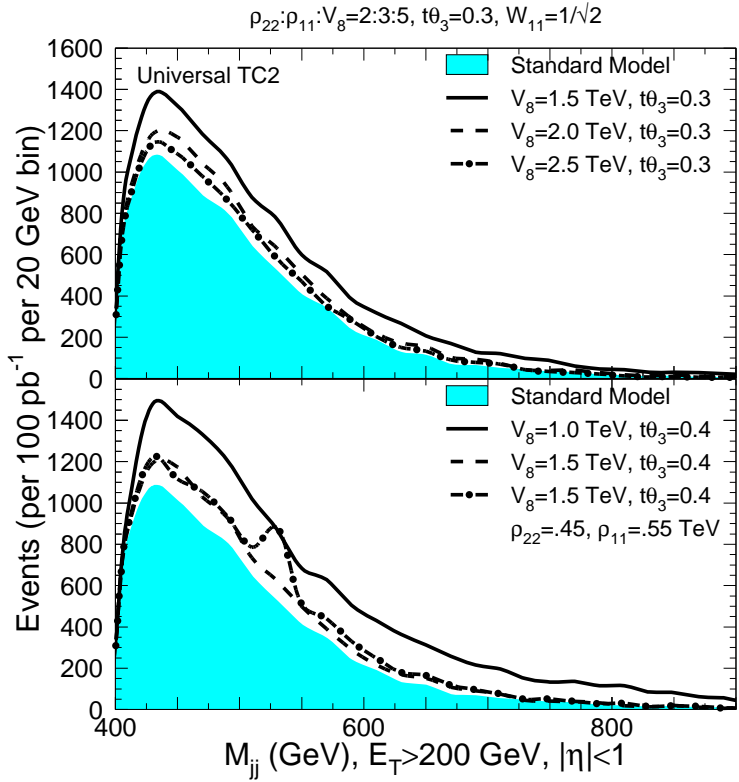,width=11cm}
    \caption{Similar to Fig.~\ref{fig:cdf}, except considering
light colorons.}
    \label{fig:bmj}
  \end{center}
\end{figure}

\leftline{{\bf Benchmark VI---Universal TC2 jets with moderate $\Mv$}}

The last benchmark studies the effect of a coloron from flavor--universal TC2
on the dijet invariant mass distributions. In Run~I, CDF~\cite{CDFjets} and
D\O~\cite{Abbott:1998yy} excluded a flavor--universal coloron with masses
less than $759-980$ GeV, depending on $\tan\thc$. We have chosen
$M_{\rho_{11,22}}=450$, 550~GeV, but they have little effect on the
distribution. The main parameters are the coloron mass and coupling. For
$\Mv=1.5\,\tev$ and $\tan\thc=0.3$ (upper panel, solid), there is a
significant enhancement over the standard model distribution. However, for
$\Mv=2\,\tev$ (dashes) and 2.5~TeV (dash--dot), there is little
distortion. Note that, for ordinary dijet production, the effect of a flavor
universal coloron below resonance is to increase the cross section (see
also~\cite{Simmons:1996fz}). This is because of the dominance of $\hat{t}$
channel coloron exchange in the QCD subprocesses, which cannot interfere
destructively.

\begin{figure}[!hbt]
  \begin{center}
    \psfig{file=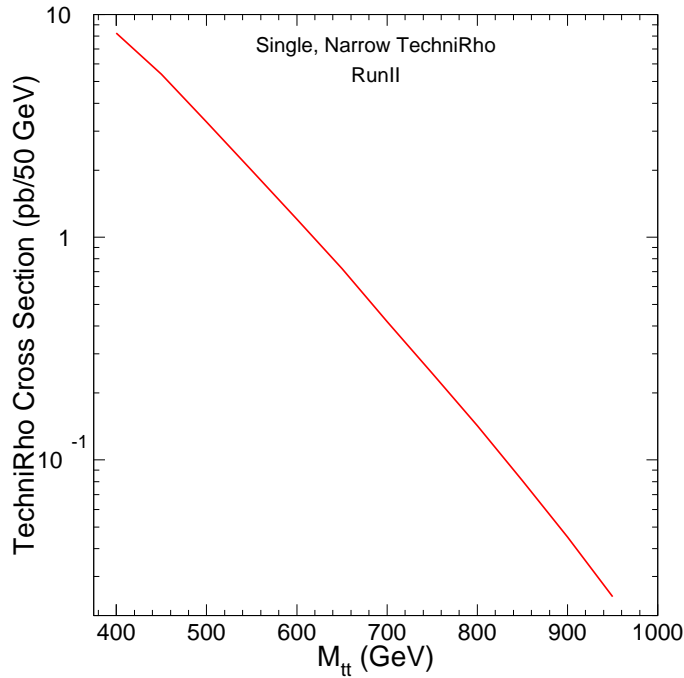,width=10cm}
    \caption{The additional cross section (in pb) 
from a narrow technirho resonance in a 50 GeV bin of the $\ol tt$ 
distribution for the Tevatron Run~II.}
    \label{fig:runii}
  \end{center}
\end{figure}

We have presented numerical results for Tevatron Run~I to indicate, crudely,
the range of TC2 parameters allowed by the data. Much more systematic studies
need to be carried out for the parameter ranges allowed by existing data and
the reach anticipated for future data. These studies should include detector
effects, efficiencies, etc. The expected improvement from Tevatron Run~II
will be statistical, resulting from the increased beam luminosity and
partonic luminosity at larger~$x$. Since the standard model $\ol tt$ cross
section increases similarly, the relative importance of colorons or
technirhos is basically unchanged. To make one more contact with Run~I
studies, we show in Fig.~\ref{fig:runii} the expected contribution of a
single, narrow technirho to the $\ol tt$ distribution (simulated by making
the other technihadrons and coloron very heavy). This case is
unrepresentative of any TC2 model.

Finally, in our examples, we focussed on TC2 effects on standard model final
states. The decays $\troct\to g \tpipr $ and $g \tpioct$ were included when
calculating the $\troct$ width, but the direct production of $\troct$
followed by the subsequent decay $\tpi$ was not. A direct search for $\troct
\ra g \tpipr \ra g \ol bb$ would be challenging. The default decay of
$\tpioct$, motivated by certain TC2 models, is to two gluons. This is
probably impossible to isolate above backgrounds. Nevertheless, a search for
$\tpioct \ra \ol bb$ (or even $\ol tt$) would be worthwhile.

\section*{8. Discussion and Conclusions}

In this paper, we improved and extended the technicolor straw man model for
testing low--scale technicolor, i.e., the notion that dynamical electroweak
symmetry breaking requires a large number of electroweak doublets of
technifermion and, therefore, that the technicolor energy scale may be only a
few hundred GeV. The improvements were made to the color--singlet sector of
the TCSM, and are largely applicable to $e^+e^-$ colliders operating below
the narrow $\tro$ and $\tom$ resonances expected at several hundred GeV. The
principal signals here are $\tpi$ production---either in association with a
$\gamma$, $W^\pm$, or $Z^0$, or in pairs---or a narrow $\ell^+\ell^-$
resonance.

We extended the TCSM by including a color--nonsinglet sector. This is based
on models of topcolor--assisted technicolor in which topcolor is broken by
technifermion condensates. This approach, summarized in Section~1, is still
the only fully--described dynamical scheme for generating the top quark's
mass. It, too, requires low--scale technicolor. Although the naturalness of
TC2---the requirement that topcolor gauge boson masses, $M_{V_8}$ and
$M_{Z'}$, not be much larger than 1~TeV---is challenged by $B$--meson mixing
measurements, there is enough room to evade those constraints that other,
independent tests of TC2 are needed. Hadron collider experiments are
particularly incisive. There, production of ordinary jet and $\ol tt$ and
$\ol bb$ pairs may reveal color octet $\troct$ and $V_8$ resonances, the
latter for subprocess energies well below $M_{V_8}$. We considered both
variants of TC2, the standard version proposed by Hill in which the coloron
couples strongly only to top and bottom quarks, and the flavor--universal TC2
advanced by Simmons and her collaborators. For a large and probable range of
TCSM parameter space, the effects of these two variants on $\ol t t$
production are striking, and strikingly different.

All of the color--singlet and nonsinglet processes discussed in this paper
are now included in the event generator {\sc Pythia}. This should provide many
happy hours of detailed studies by experimental collaborations at the
Tevatron and the LHC, as well as refined reanalysis of data from LEP and
opportunities for those yearning for studies of nonsupersymmetric physics in
the TeV region of a linear $e^+e^-$ collider. An application to LEP physics
was presented in Ref.~\cite{Lane:2002wb}. To illuminate our discussion for
hadron colliders, we considered several benchmark parameter sets and studied
their effect on observables for Tevatron Run~I conditions. Our parton--level
studies with {\sc Pythia} are qualitative. A quantitative discussion of jet
or heavy quark observables requires detailed simulations, including features
of real detectors and running conditions. We are encouraged that, where a
comparison is possible, our numerical results are in reasonably good
agreement with published Run~I limits on colorons and technirhos.

Top quark pair production at the Tevatron is an ideal probe for topcolor, TC2
and other new dynamics associated with producing $m_t$, since the final state
is distinctive and it is produced through strong interactions dominated by
valence quark annihilation. There is great sensitivity to the coloron even
far below resonance. Furthermore, signals for the technirhos, which are
expected to be lighter than the coloron, are sensitive to the coloron mass
through mixing effects. For standard TC2, the coloron interferes
constructively with the gluon in $s$-channel processes, and enhances the
production cross section, fairly independently of the $SU(3)$ mixing
parameter, $\tan\thc$. Without including relatively light $\troct$, the Run~I
lower bound on the standard TC2 coloron mass is $\Mv > 1$--2~TeV. On the
other hand, there is a significant reduction of the cross section for
flavor--universal TC2. This effect is further enhanced by the direct
appearance of $\tan\thc$ in the effective couplings between light and heavy
quarks. The Run~I coloron bound is $\Mv > 3$--4~TeV, depending on
$\tan\thc$. Since $\troct$ are expected to be lighter than the coloron, their
appearance as narrow peaks in heavy or light quark pair production places
more stringent bounds on the coloron mass in either model.

\section*{Acknowledgements}

The model described here will be available in {\sc Pythia v6.211}.

We are grateful for inspiration, advice, and encouragement from Guennadi
Borissov, Sekhar Chivukula, Estia Eichten, Robert Harris, Chris Hill, Andrei
Kounine, Meenakshi Narain, Stephen Parke, Francois Richard, Markus
Schumacher, Elizabeth Simmons, John Womersley and other members of the
``Strong Dynamics for Run~II Workshop'' at Fermilab. KL's research was
supported in part by the Fermilab Theory Group Frontier Fellowship program
and by the Department of Energy under Grant~No.~DE--FG02--91ER40676. FNAL is
operated by Universities Research Association Inc., under contract
DE-AC02-76CH03000.

\newpage

\section*{Appendix A: Default Values for Parameters}

The suggested default values of the parameters used in this note are listed
in Table~3.
{\begin{table}[!ht]{
\begin{tabular}{|c|c|c|c|}
\hline
Parameter& Default Value&  Parameter&  Default Value
\\
\hline\hline
$\Ntc$& 4 & $\atro(\Ntc/3)$ & 2.91  \\
$\sin\chi$  & $\third$ & $\sin\chipr$  & $1/\sqrt{6}$\\
$Q_U$  & $\fourthirds$ &$Q_D = Q_U-1$  & $\third$ \\
%%%$C_{1c,b}$& 1 & $C_{1c}$& 1\\
$C_{1\tau,c,b}$& 1 & $C_{1t}$& $m_b/m_t$\\
$C^2_{1g}$& $\tx{4\over{3}}$ & $\vert\epsilon_{\rho\omega}\vert$ & 0.05 \\
\hline
$F_T = F_\pi \sin\chi$  & $82\,\gev$ & $M_{V,A}$ & $200\,\gev$ \\
$M_{\tropm,\tom,\troz}$ & $210\,\gev$ & $M_{\tpipm,\tpiz,\tpipr}$ & $110\,\gev$ \\
%$M_{\tpiz}$ & $110\,\gev$ & $M_{\tpipr}$ & $110\,\gev$ \\
%%$M_{V}$ & $200\,\gev$ & $M_{A}$ & $200\,\gev$ \\
\hline
$\sin\chi''$ &$1/\sqrt{2}$ & $\tan\thc$ & $\sqrt{0.08}$ \\
$C_{8s,c,b,t}$ & 0 & $C^2_{8g}$ & $\tx{5\over{3}}$ \\
%& $C_{8b}$ & 0 \\
%$C_{8c}$ & 0 & $C_{8s}$ & 0 \\
%%$g_{u,d,c,s}$ & $-g_C \tan\thc$ & $g_{t,b}$ & $g_C \cot\thc$ \\
%%$g_{u,d,c,s}$ & $g_C \cot\thc$ & $g_{t,b}$ & $g_C \cot\thc$ \\
$M_{\tpioct}$ & $250\,\gev$ & $\Mv$ & $500\,\gev$ \\
$M_{\rho_{11}}$ & $400\,\gev$ & $M_{\rho_{22}}$ & $300\,\gev$ \\
$M_{\rho_{12,12'}}$ & $350\,\gev$ &
$M_8,M'_8$ & $250\,\gev$ \\
%%& $M'_8$ & $250\,\gev$ \\
$W_{L11}^{U,D}$ & $1/\sqrt{2}$ & $\phi_{U,D}$ & $0$  \\
%$W_{L11}^D$ & $1/\sqrt{2}$ & $\phi_D$ & $0$  \\
\hline\hline
\end{tabular}}
\caption{Default values for parameters in the
  Technicolor Straw Man Model. As in Refs.~\cite{tcsm,elw}, the technipion
  decay constant is $F_T = F_\pi \sin\chi$ with a default value of 82~GeV,
  and the $\tro \ra \tpi\tpi$ coupling is $\atro = 2.91(3/\Ntc)$. 
The default model is standard TC2~\cite{tctwohill}. We have set
$W_{R}^{U,D}= 1$ and used Eq.~(\ref{eq:unitarity}) for the matrix elements
$W_{Lij}^{U,D}$.}
\end{table}}

\vfil\eject


\begin{thebibliography}{99}
%
\bibitem{tcsm} K.~Lane, Phys.~Rev.~{\bf D60}, 075007 (1999) [\myref{hep-ph/9903369}].
%
%
\bibitem{ehs} L.~Randall and E.~H.~Simmons, Nucl.~Phys.~{\bf B380}, 3
  (1992);\\
G.~Rupak and E.~H.~Simmons, Phys.~Lett.~{\bf B362}, 155 (1995)
[\myref{hep-ph/9507438}]; \\
K.~R.~Lynch and E.~H.~Simmons, Phys.~Rev.~{\bf D64}, 035008 (2001)
[\myref{hep-ph/0012256}].
%
%
\bibitem{Lane:2002wb}
K.~Lane, {\it et al.},
%, K.~R.~Lynch, S.~Mrenna and E.~H.~Simmons,
%``Resonant and non-resonant effects in photon technipion production at  lepton colliders,''
Phys.~Rev.~{\bf D66}, 015001 (2002) [\myref{hep-ph/0203065}].
%
\bibitem{Mrenna:1999ks}
S.~Mrenna,
%``Technihadron production and decay at LEP2,''
Phys.\ Lett.\ B {\bf 461}, 352 (1999)
[\myref{hep-ph/9907201}].
%
\bibitem{l3} The L3 Collaboration, L3 Physics Note 2428.
%
%
\bibitem{delphi} J.~Abdallah, {\it et~al.}, The DELPHI Collaboration,
Eur.~Phys.~J.~{\bf C22}, 17 (2001) [\myref{hep-ex/0110056}].
%
%
\bibitem{opal}The OPAL Collaboration, {\it Searches for Technicolor with the
    OPAL Detector in $e^+e^-$ Collisions at the Highest LEP Energies},
    (2001), OPAL Physics Note PN485.
%
%
\bibitem{pythia} T.~Sj\"ostrand, {\it et al.}, Comput.~Phys.~Commun.~{\bf 135},
238, (2001) [\myref{hep-ph/0010017}];\\
T.~Sj\"ostrand, L.~Lonnblad, and S.~Mrenna, (2001) [\myref{hep-ph/0108264}].
%
%
\bibitem{tc} S.~Weinberg, Phys.~Rev.~{\bf D19}, 1277 (1979);\\
L.~Susskind, Phys.~Rev.~{\bf D20}, 2619 (1979).
%
%
\bibitem{etc} E.~Eichten and K.~Lane, Phys.~Lett.~{\bf B90}, 125 (1980).
%
%
\bibitem{wtc}B.~Holdom, Phys.~Rev.~{\bf D24}, 1441 (1981);\\
Phys.~Lett.~{\bf 150B}, 301 (1985);\\
T.~Appelquist, D.~Karabali and L.~C.~R. Wijewardhana,
Phys. Rev. Lett.~{\bf 57}, 957 (1986);\\
T.~Appelquist and L.~C.~R.~Wijewardhana, Phys.~Rev.~{\bf D36}, 568
(1987);\\
K.~Yamawaki, M.~Bando and K.~Matumoto, Phys.~Rev.~Lett.~{\bf 56}, 1335
(1986);\\
T.~Akiba and T.~Yanagida, Phys.~Lett.~{\bf 169B}, 432 (1986).
%
%
\bibitem{topcref}
C.~T. Hill, Phys.~Lett.~{\bf 266B}, 419 (1991);\\
S.~P.~Martin, Phys.~Rev.~{\bf D45}, 4283 (1992);\\
{\it ibid}~{\bf D46}, 2197 (1992); Nucl.~Phys.~{\bf B398}, 359 (1993);\\
M.~Lindner and D.~Ross, Nucl.~Phys.~{\bf  B370}, 30 (1992);\\
R.~B\"{o}nisch, Phys.~Lett.~{\bf 268B}, 394 (1991);\\
C.~T.~Hill, D.~Kennedy, T.~Onogi, H.~L.~Yu, Phys.~Rev.~{\bf D47}, 2940 
(1993).
%
%
\bibitem{tctwohill}C.~T.~Hill, Phys.~Lett.~{\bf 345B}, 483 (1995)
[\myref{hep-ph/941142}].
%
%
\bibitem{multi}K.~Lane and E.~Eichten, Phys. Lett. {\bf B222}, 274 (1989).
%
%
\bibitem{elw} E.~Eichten and K.~Lane, Phys.~Lett.~{\bf B388}, 803 (1996)
  [\myref{hep-ph/9607213}];\\
E.~Eichten, K.~Lane and J.~Womersley, Phys.~Lett.~{\bf B405}, 305 (1997)
  [\myref{hep-ph/9704455}];\\
E.~Eichten, K.~Lane and J.~Womersley, Phys.~Rev.~Lett.~{\bf 80}, 5489
  (1998) [\myref{hep-ph/9802368}].
%
%
\bibitem{CDFWtpi} T.~Affolder, {\it et al.}, The CDF Collaboration,
  Phys.~Rev.~Lett.~{\bf 84}, 1110 (2000).
%
%
\bibitem{CDFgtpi} F.~Abe, {\it et al.}, The CDF Collaboration,
  Phys.~Rev.~Lett.~{\bf 83}, 3124 (1999) [\myref{hep-ex/9810031}].
%
%
\bibitem{CDFjets} F.~Abe, {\it et al.}, The CDF Collaboration,
Phys.~Rev.~{\bf D55}, 5263 (1997) [\myref{hep-ex/9702004}].
%
%
\bibitem{Hoffman:mk} F. Abe {\it et al.},
The CDF~Collaboration, 
  Phys.~Rev.~Lett.~{\bf 82}, 2038 (1999) [\myref{hep-ex/9809022}].
%K.~D.~Hoffman,
%``The Search For Physics Beyond The Standard Model In The B Anti-B Spectrum
%Observed In Tevatron Proton - Anti-Proton Collisions,''
%FERMILAB-THESIS-1998-38.
%
\bibitem{CDFLQ} F.~Abe, {\it et al.}, The CDF Collaboration,
Phys.~Rev.~Lett.~{\bf 82}, 3206 (1999).
%
%
\bibitem{ATLAS} Atlas Physics Technical Design Report, Chapter~21 (1999),
available at
\myref{\verb+http://atlasinfo.cern.ch/Atlas/GROUPS/PHYSICS/TDR/access.html+}.
%
%
\bibitem{tcsmrates}K.~Lane, {\it Technicolor Production and Decay Rates in
    the Technicolor Straw Man Model}, [\myref{hep-ph/9903372}], Boston University
    Preprint BUHEP-99-5, March 1999. Note that the present paper supersedes
    the production, but not the decay, rates compiled in this paper.


%
%
\bibitem{tctwokl} K.~Lane, Phys.~Rev.~{\bf D54}, 2204 (1996) [\myref{hep-ph/9602221}];\\
K.~Lane, Phys.~Lett.~{\bf B433}, 96 (1998) [\myref{hep-ph/9805254}].
%
%
\bibitem{ehlq} E.~Eichten, I.~Hinchliffe, K.~Lane and C.~Quigg,
  Rev.~Mod.~Phys~{\bf 56}, 579 (1984); Phys.~Rev.~{\bf DD34}, 1547 (1986).
%
%
\bibitem{multiklrm} K.~Lane and M.~V.~Ramana, Phys.~Rev.~{\bf D44}, 2678
  (1991).
%
%
\bibitem{ctzprime} R.~S.~Chivukula and J.~Terning, Phys.~Lett.~{\bf B385},
  209 (1996) [\myref{hep-ph/9606233}].
%
%
\bibitem{trzprime} T.~Rador, Phys.~Rev.~{\bf D59}, 095012 (1999) [\myref{hep-ph/9810252}].

%
%
\bibitem{ccs}R.~S.~Chivukula, A.~G.~Cohen and E.~H.~Simmons, 
Phys.~Lett.~{\bf B380}, 92 (1996) [\myref{hep-ph/9603311}];\\
M.~Popovic and E.~H.~Simmons, Phys.~Rev.~{\bf D58}, 095007
(1998) [\myref{hep-ph/9806287}].
%
%
\bibitem{blt} G.~Burdman, K.~Lane, and T.~Rador, Phys.~Lett.~{\bf B514}, 41
(2001) [\myref{hep-ph/0012073}].
%
%
\bibitem{ehsbbmix} E.~H.~Simmons, Phys.~Lett.~{\bf 526}, 365 (2002) [\myref{hep-ph/0111032}].
%
%
\bibitem{ibehs} I.~Bertram and E.~H.~Simmons, Phys.~Lett.~{\bf B443},
347 (1998) [\myref{hep-ph/9809472}].
%
%
\bibitem{vacalign} K.~Lane, T.~Rador and E.~Eichten,  Phys.~Rev.~{\bf D62},
  015005 (2000) [\myref{hep-ph/0001056}];\\
K.~Lane, {\it Strong and Weak CP Violation in Technicolor}, Invited talk at
  the Eighth International Symposium on Particles, Strings and
  Cosmology---PASCOS 2001, University of North Carolina, Chapel Hill, NC,
  April 10--15, 2001 [\myref{hep-ph/0106328}].
%
%
\bibitem{cdt} R.~S.~Chivukula, B.~Dobrescu and J.~Terning, 
Phys.~Lett.~{\bf B353}, 289 (1995) [\myref{hep-ph/9503203}].
%
%
\bibitem{tctwoklee} K.~Lane and E.~Eichten, Phys.~Lett.~{\bf B352}, 382
(1995) [\myref{hep-ph/9503433}].
%
%
\bibitem{etat} E.~Eichten and K.~Lane, Phys.~Lett.~{\bf B327}, 129 (1994)
  [\myref{hep-ph/9401236}].
%
%
\bibitem{ZR} A.~Zerwekh and R.~Rosenfeld, Phys.~Lett.~{\bf B503}, 325 (2001)
[\myref{hep-ph/0103159}].
%
%
\bibitem{cgs} R.~S.~Chivukula, A.~Grant and E.~H.~Simmons, Phys.~Lett.~{\bf
B521}, 239 (2001) [\myref{hep-ph/0109029}].
%
%
\bibitem{cthsjp} C.~T.~Hill and S.~Parke, Phys.~Rev.~{\bf D49}, 4454 (1994)
[\myref{hep-ph/9312324}].
%
%
\bibitem{kltop} K.~D.~Lane, Phys.~Rev.~{\bf D52}, 1546 (1995) [\myref{hep-ph/9501260}].
%
%
\bibitem{cdftop} 
T. Affolder {\it et al.}, The CDF~Collaboration,
Phys. Rev. Lett. 85, 2062 (2000) [\myref{hep-ex/0003005}].
%
\bibitem{d0top} B. Abbott {\it et al.},
The D\O~Collaboration, 
Phys. Rev. {\bf D58}, 052001 (1998) [\myref{hep-ex/9801025}].


\bibitem{Simmons:1996fz} E.~H.~Simmons,
%``Coloron phenomenology,''
Phys.\ Rev.\ D {\bf 55}, 1678 (1997)
[\myref{hep-ph/9608269}].
%%CITATION = HEP-PH 9608269;%%
%
%\cite{Abbott:1998yy}
\bibitem{Abbott:1998yy}
B.~Abbott {\it et al.}, The D\O\ Collaboration, 
%``Coloron limits using the D0 dijet angular distribution,''
[\myref{hep-ex/9809009}].
%%CITATION = HEP-EX 9809009;%%

\end{thebibliography}
\end{document}